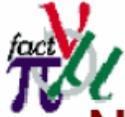

International scoping study of a future Neutrino Factory and super-beam facility

RAL-TR-2007-23

# Accelerator design concept for future neutrino facilities

J. S. Berg[a], A. Blondel[b], A. Bogacz[c], S. Brooks[d], J.-E. Campagne[e], D. Caspar[f], C. Cevata[g], P. Chimenti[h], J. Cobb[i], M. Dracos[j], R. Edgecock[d], I. Efthymiopoulos[k], A. Fabich[k], R. Fernow[a], F. Filthaut[l], J. Gallardo[a], R. Garoby[k], S. Geer[m], F. Gerigk[k], G. Hanson[n], R. Johnson[o], C. Johnstone[m], D. Kaplan[p], E. Keil[k,q], H. Kirk[a], A. Klier[n], A. Kurup[r], J. Lettry[k], K. Long[r], S. Machida[d], K. McDonald[s], F. Méot[t], Y. Mori[u], D. Neuffer[m], V. Palladino[v], R. Palmer[a], K. Paul[o], A. Poklonskiy[w], M. Popovic[m], C. Prior[i], G. Rees[d], C. Rossi[k], T. Rovelli[x], R. Sandström[b], R. Sevior[y], P. Sievers[k,q], N. Simos[a], Y. Torun[m,p], M. Vretenar[k], K. Yoshimura[z], and M. S. Zisman[aa,ab]

[a]*Brookhaven National Laboratory, Upton, Long Island, NY, USA*
[b]*University of Geneva, Geneva, Switzerland*
[c]*Thomas Jefferson National Accelerator Facility, Newport News, VA, USA*
[d]*Rutherford Appleton Laboratory, UK*
[e]*LAL, University Paris-Sud, IN2P3/CNRS, Orsay, France*
[f]*University of California–Irvine, Irvine, CA, USA*
[g]*CEA, France*
[h]*University of Trieste and INFN, Trieste, Italy*
[i]*University of Oxford, Oxford, UK*
[j]*Institut de Recherches Subatomiques, Université Louis Pasteur, Strasbourg, France*
[k]*CERN, Geneva, Switzerland*
[l]*NIKHEF, Amsterdam, The Netherlands*
[m]*Fermi National Accelerator Laboratory, Batavia, IL, USA*
[n]*University of California–Riverside, Riverside, CA, USA*
[o]*Muons, Inc., Batavia, IL, USA*
[p]*Illinois Institute of Technology, Chicago, IL, USA*
[q]*Retired*
[r]*Imperial College London, London, UK*
[s]*Princeton University, Princeton, NJ, USA*
[t]*CEA and IN2P3, LPSC, Grenoble, France*
[u]*RRI, Kyoto University, Osaka, Japan*
[v]*INFN-Napoli, Naples, Italy*
[w]*Michigan State University, East Lansing, MI, USA*
[x]*INFN-Bologna, Bologna, Italy*
[y]*Lancaster University, Lancaster, UK*
[z]*KEK, Tsukuba, Japan*
[aa]*Lawrence Berkeley National Laboratory, Berkeley, CA, USA*
[ab]*Editor*

The ISS Accelerator Working Group

September 10, 2008

# 1. Introduction

This document summarizes the findings of the Accelerator Working Group (AWG) of the International Scoping Study (ISS) of a Future Neutrino Factory and Superbeam Facility. The work of the group took place at three plenary meetings along with three workshops, and an oral summary report was presented at the NuFact06 workshop held at UC-Irvine in August, 2006. The goal was to reach consensus on a baseline design for a Neutrino Factory complex. One aspect of this endeavor was to examine critically the advantages and disadvantages of the various Neutrino Factory schemes that have been proposed in recent years. This comparison is discussed in Section 4.

The activities of the group were coordinated by an Accelerator Council, whose members are listed in Table 1. Initially, a series of issues and tasks was identified for each of the various subsystems that comprise a Neutrino Factory. Over the course of the one-year study, an attempt was made to address these. The status of this work is summarized here. In addition, a list of required R&D activities was developed as a guide to future effort; its main items are summarized in Section 7.

## *1.1 Issues Addressed During the ISS*

Proton Driver
- What is the optimum beam energy (which depends to some degree on the choice of target material)?
- What is the optimum repetition rate?
- What is the optimum bunch length?
- Is there a preferred hardware configuration (e.g., linac, synchrotron, FFAG ring,…)?

Target, Capture and Decay
- What is the optimum target material (high or low *Z*)?
- What are the target limitations on proton beam parameters at 4 MW (bunch intensity, bunch length, pulse duration, repetition rate)?
- How do Superbeam and Neutrino Factory requirements compare?

Front End
- Compare existing schemes, both with and without cooling
- Consider effects of reduced operating specifications
- Examine trade-offs between cooling and downstream acceptance

Table 1. Accelerator Council members.

| | |
|---|---|
| R. Fernow | BNL |
| R. Garoby | CERN |
| Y. Mori | Kyoto University |
| R. Palmer | BNL |
| C. Prior | Oxford/ASTeC |
| M. Zisman, convener | LBNL |

Acceleration
- Compare alternative schemes (linac, RLA, FFAG) on an equal footing
- Examine implications of increased acceptance
- Study dynamics and matching with errors

Decay Ring
- Design implications of final energy (20 vs. 50 GeV)
- Implications of keeping both muon signs
- Implications of two simultaneous baselines
- Optics requirements *vs.* emittance

## *1.2 Organization of Report*

In what follows, we discuss parameters and concepts for the proton driver (Section 2), the target (Section 3), the front end (Section 4, which also includes a performance comparison of the various designs that have been developed in recent years), the acceleration system (Section 5) and the decay ring (Section 6). Section 7 gives a description of the key elements of the R&D plan, many of which are already well under way. Section 8 provides a brief summary of what we have learned.

## 2. Proton Driver

### *2.1 Introduction*
Many factors influence the specifications for the proton driver. Among these are:

- the required production of $\approx 10^{21}$ neutrinos per year
- muon yields as a function of the proton energy
- muon yields as a function of the target material
- heating and stress levels for the target material
- muon capture as a function of proton bunch length
- maximum acceptable duration of proton pulses on the target
- peak beam loading levels in the $\mu^\pm$ accelerators
- bunch train stacking in the $\mu^+$ and $\mu^-$ decay rings

After considering all of these, the proton driver specifications for the ISS were set as indicated in Table 2. As can be seen, there are differing—and incompatible—requirements for liquid-Hg and solid metallic targets. In this report, as discussed in Section 3, our baseline target choice is the liquid-Hg jet, so our efforts have focused mainly on those parameters in the designs described here. A solid target could be accommodated with some changes in design parameters.

### *2.2 Proton Driver Options*
A number of options were considered for a 4 MW, 50 Hz proton driver. These include:

- an H⁻ linac with a 50-Hz booster RCS and a 50-Hz non-scaling, non-linear, fixed-field alternating gradient (NFFAG) driver ring
- an H⁻ linac with pairs of 50 Hz booster and 25 Hz driver synchrotrons (RCS)

Table 2. Proton driver requirements.

| Parameter | Value |
|---|---|
| Average beam power (MW) | 4 |
| Pulse repetition frequency (Hz) | 50 |
| Proton energy (GeV) | $10 \pm 5$ |
| Proton rms bunch length (ns) | $2 \pm 1$ |
| No. of proton bunches | 3 or 5 |
| Sequential extraction delay (µs) | $\geq 17$ |
| Pulse duration, liquid-Hg target (µs) | $\leq 40$ |
| Pulse duration, solid target (ms) | $\geq 20$ |

- an H$^-$ linac with a chain of three non-scaling FFAG rings in series
- an H$^-$ linac with two slower cycling synchrotrons and two holding rings
- a full energy H$^-$ linac with an accumulator and bunch compression ring(s)

Of these options, the most advanced design is for the first, and this is described here to give a sense of what a workable system must include. In the same spirit, the last option, using an energy at the low end of the desired range, is also briefly described in Section 2.5.

As can be seen in Table 2, we have chosen to specify proton driver performance requirements rather than a specific implementation. This is because there is not a unique solution to providing our requirements, and any solution that does so would be acceptable. The actual choice at a particular host site will undoubtedly be dictated by many factors, including cost, local expertise, and other possible uses of the proton driver complex. Thus, the examples here should be taken as indicative of possible approaches rather than as endorsements for a particular approach.

## *2.3 Proton and Muon Bunch Train Patterns*

Proton bunch compression occurs in each 50 Hz cycle, with five bunches preferred.[1] To keep the pulse duration below 40 µs for the Hg-jet target, however, only three bunches can be used. Each proton bunch creates pions in the target, and these decay to give a single µ$^\pm$ bunch, which is then transformed to a train of interleaved 80 µ$^+$ and 80 µ$^-$ bunches in a bunch rotation scheme [1]. A schematic diagram of the proposed bunch patterns is given in Fig. 1.

For a uniform pattern in both decay rings, the three (397.5 ns) bunch trains are separated by 993.8 ns time gaps. If the rings are in a single tunnel, or if both beams are stored in a single ring, the µ$^+$ trains in one ring are interleaved in time with the µ$^-$ trains in the other. In the example shown in Fig. 1, the time gaps between the µ$^+$ and µ$^-$ bunch trains are 298 ns for the three trains. An RF system is needed in each ring to contain the 201.25 MHz bunches and preserve the gaps.

---

[1]The use of multi-bunch trains at 50 Hz is a change made during the study from the original single, 15-Hz train. The change was made to ease the production of the $2 \pm 1$ ns (rms) proton bunches, and to reduce the heavy beam loading in the µ$^\pm$ accelerators.

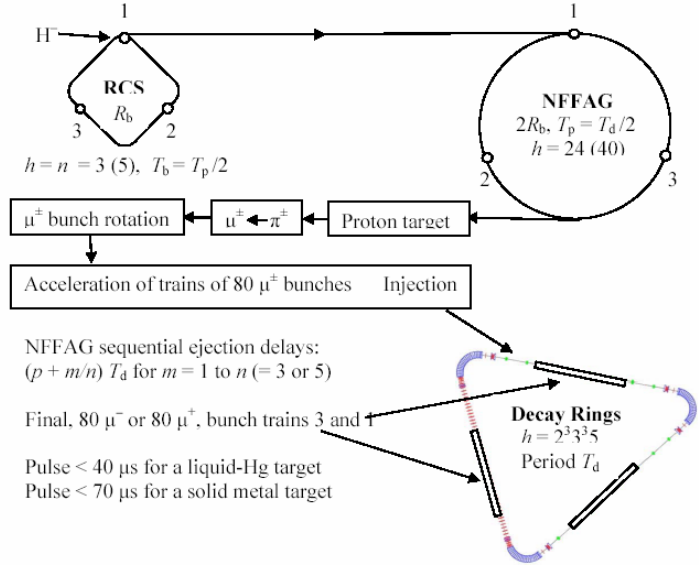

Fig. 1. Bunch patterns ($n = 3$) for the proton driver and decay rings. This example assumes triangular decay rings.

### *2.4 10 GeV, NFFAG Proton Driver Complex*

Prior to the ISS, a 50 Hz, 10 GeV, 4 MW proton driver [2] was designed at RAL. During the study, the design was modified for a three (five), bunch compatibility between booster, driver and 20–50 GeV muon decay rings.

The booster energy range is 0.2–3 GeV, and the proton driver energy range is 3–10 GeV. The driver is a new type of FFAG accelerator that uses a non-isochronous, non-linear, and non-scaling cell focusing structure (which we denote as "NFFAG"). Either three or five proton bunches may be used with the design.

2.4.1 H$^-$ Injector Linac

A 90 MeV injector linac design at RAL [3] is extended to 200 MeV by adding a 110 MeV side coupled linac. The frequency is 324 MHz[2], and can be provided by commercially available klystrons. The pulse repetition frequency is 50 Hz, the peak current ~30 mA, the pulse duration ~400 μs, and the duty cycle after chopping ~70%.

2.4.2 Linac-to-Booster Transfer Line

An achromatic section of beam line is used between the linac and the booster ring, both for collimation and for diagnostic purposes. Twelve combined-function magnets, arranged as four triplet cells, form the achromat. Eight of the magnets have +45° bends and four have –45° for a net bend of 180°. The length of the line is 41.6 m, and the peak dispersion and normalized dispersion functions at the central symmetry point are 14.16 m and 5.1 m$^{1/2}$, respectively. Achromaticity is obtained by using mirror symmetry about the bend center, and choosing a π horizontal betatron phase shift, including space-charge, for the first two and the last two triplet cells. The negative bends are in the first and the last magnet of each π section.

---

[2] The possibility of tripling the frequency of the linac sections beyond 90 MeV will be considered as the design progresses.

Upstream of the achromat is a beam line with horizontal beam-loss collimators and cavities for momentum spread reduction and correction. In the achromat are more collimators and bunchers. There is vertical collimation for both beam edges at the four FDF triplet centers, and momentum edge-collimation at two places. The main momentum collimation is at the high dispersion center point, with pre-collimation in the preceding triplet. Stripping foils are used rather than conventional collimators. The combined-function magnets used for the line are C-magnets, open on the outer radius for easy exit of the stripped ions to shielded beam dumps. Bunchers at the one-quarter and three-quarters arc positions restore the upright orientation of the momentum phase space at the arc center and end.

### 2.4.3  50 Hz Booster Synchrotron

A 50 Hz synchrotron is used for a booster. The injection scheme dominates the lattice design, which is based on that for a European Spallation Source (ESS) [4]. The lattice comprises four superperiods, each having seven triplet cells. In each superperiod, three of the seven cells form a 90° arc and four cells form a dispersion-free straight section. All sixteen of the straight cells have a 10.6 m free length. An 8°, 5.4 m, dipole is in the center cell of each arc, and adjacent cells each contain two, 20.5°, 4.15 m main dipoles. All of the injection elements are located between the two central triplets of one of the arcs, thus providing a fully separated injection scheme. The booster layout is shown in Fig. 2.

Quadrupole gradients for the symmetrical triplets on each side of a central dipole are adjusted for zero dispersion straight sections. Two other quadrupole types vary the betatron tunes and a fifth type adjusts the normalized dispersion at the arc center. Because the ring is fast cycling, we choose a common gradient for all 84 quadrupoles. The five different quadrupole types are distinguished by length differences, simplifying the quadrupole power supply requirements. Parameters are given in Table 3 and Twiss parameters are shown in Fig. 3.

Injection makes use of an H¯ stripping foil located in the center of an 8° arc dipole. A normalized dispersion of ~1.9 $m^{1/2}$ allows horizontal phase-space beam painting through related RF steering and momentum ramping of the input beam. An injection chicane is not needed, nor are horizontal painting magnets or an injection septum magnet. Longitudinal painting is not as easy as injecting into a dispersion-free region, but earlier space-charge tracking studies of injection [5] proved satisfactory. Two sets of symmetrical steering magnets, one pair on each side of the 8° dipole, provide the vertical beam painting. Either an anti-correlated or a correlated transverse beam distribution can be obtained. The betatron tune depression at 200 MeV is ~ 0.25 for an assumed two-dimensional elliptical density distribution.

The dipole magnets are at low-beta positions in the triplet cells to keep their stored energy low, which is important in the design of the main-magnet power supplies. During acceleration, the fields in the 8° dipoles cycle between 0.055 and 0.33 T, and those in the main 20.5° dipoles, between 0.19 and 1.1 T. Gradients in the quadrupoles track between 1.0 and 5.9 T m$^{-1}$. Designs for the magnets and magnet power supplies have not yet been done. In addition to the main magnets, the use of 32 vertical and 32 horizontal steering magnets, and the same number of horizontal and vertical trim quadrupoles, is envisioned. Spaces are also reserved in the ring for sextupole magnets.

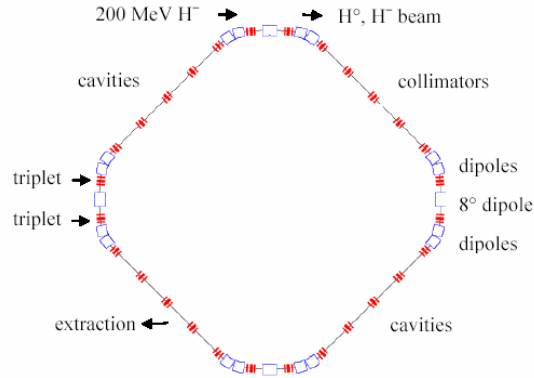

Fig. 2. Layout of 50 Hz, 3 GeV booster synchrotron.

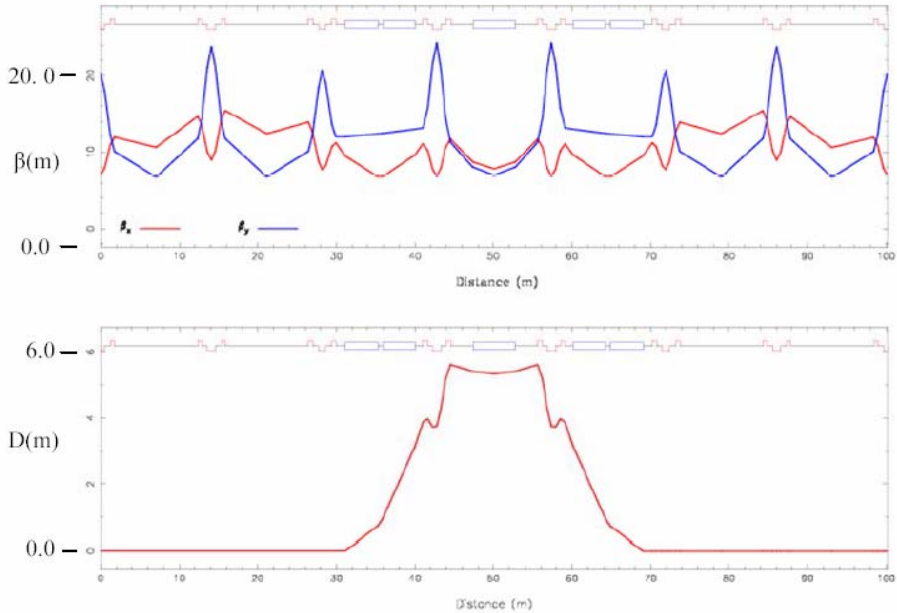

Fig. 3. (upper) Betatron functions for the 400.8 m booster synchrotron; (lower) booster dispersion function.

At a 50-Hz cycle rate, ceramic vacuum chambers with contoured RF shields, based on designs used at the ISIS synchrotron [6], are required for the main and the correction magnets. The injection magnet requires a specialized ceramic chamber having a central T-section for mounting and removal of the H⁻ stripping foil and its electron collector. This is a difficult mechanical design area because of the need to reduce eddy current effects. A standard metal and ceramic uhv vacuum system is proposed, using demountable joints with tapered flanges, band clamps and metal seals for quick connection and removal. Oil-free turbomolecular roughing pumps are used to reduce the pressure to $<10^{-6}$ mbar, after which they are removed from the ring. Ion pumps are then used at pressures of $<10^{-7}$ mbar; these also provide pressure monitoring around the ring.

Table 3. Main parameters for booster synchrotron.

| Parameter | Value | Parameter | Value |
|---|---|---|---|
| Circumference (m) | 400.8 | No. protons per cycle | $5 \times 10^{13}$ |
| Betatron tunes ($Q_y$, $Q_x$) | 6.38, 6.30 | Beam power @ 3 GeV (MW) | 1.2 |
| $\gamma_{transition}$ | 6.57 | RF straight sections (m) | $11 \times 10.6$ |
| $L_{cell}$ (m) | 14.6, 14.1 | Freq. for $h = n = 5$ (MHz) | 2.12–3.63 |
| Acceptance ($\pi$ mm-mrad) | 400 | $h = 5$ bunch area (eV-s) | 0.66 |
| Max. $\varepsilon_x$, $\varepsilon_y$ ($\pi$ mm-mrad) | 175 | $V_{RF}$ @ 3 GeV, $\eta_{SC} < 0.4$ (MV) | 0.42 |
| $L_{QD}$ | 1.07, 0.93, 1.01 | $V_{RF}$ @ 5 ms for $\phi_s = 48°$ (MV) | 0.90 |
| $L_{QF}$ | 0.51, 0.62 | Freq. for $h = n = 3$ (MHz) | 1.27–2.18 |
| Quadrupole gradient (T/m) | 0.995–5.91 | $h = 3$ bunch area (eV-s) | 1.1 |
| Quadrupole bore radius (mm) | 110 | $V_{RF}$ @ 3 GeV, $\eta_{SC} < 0.4$ (MV) | 0.25 |
| | | $V_{RF}$ @ 5 ms for $\phi_s = 52°$ (MV) | 0.85 |
| Long dipole (20.5°) | | Long dipole (8°) | |
|   No. | 16 |   No. | 4 |
|   Length (m) | 4.15 |   Length (m) | 5.45 |
|   Bend radius (m) | 11.6 |   Bend radius (m) | 38.99 |
|   Field (T) | 0.19–1.10 |   Field (T) | 0.055–0.33 |
|   h,v good field region (mm) | 160, 175 |   h,v good field region (mm) | 130, 205 |

There is a dedicated region for beam loss collimation in one ring superperiod. A momentum collimator protects ring components from longitudinal beam loss, and primary and secondary betatron collectors are used to localize the transverse beam losses in both planes.

As the 1.2 MW beam power of the booster is higher than in any existing RCS, more than 100 m is provided for the RF acceleration system. The fields used for the proton acceleration give an adiabatic bunch compression in both booster and driver. The booster uses harmonic $h = 5$ (2.117–3.632 MHz) for five 0.66 eV-s bunches, or $h = 3$ (1.270–2.179 MHz) for three 1.1 eV-s bunches. Due to the bunch compression in the booster, the proton driver can make use of eight times higher harmonic numbers. Thus, the driver has $h = 40$ (14.529–14.907 MHz) or $h = 24$ (8.718–8.944 MHz). For the case of five bunches, the booster needs RF voltages per turn of 0.9 MV for the acceleration and 0.42 MV at 3 GeV, while the driver requires 1.18 MV for an adiabatic bunch compression to ~2.1 ns rms at 10 GeV. In the three-bunch case, the corresponding parameters are 0.85, 0.25, and 1.30 MV per turn for 3.0 ns rms bunches. A hardware issue to be resolved is the choice between Finemet and ferrite for frequency tuning of the RF cavities. Some further compression is envisaged by adding higher harmonic cavities in the driver ring. A possible scheme based on using pairs of detuned cavities, at an RF phase shift of π apart to cancel reactive beam loading components, will be evaluated. Use of a multi-pulse kicker and septum extraction system, and conventional diagnostics, is assumed.

### 2.4.4  50 Hz Proton Driver

A proton driver based on an NFFAG ring has several advantages:

- It can have a high duty cycle, and thus lower RF accelerating fields.
- Adiabatic compression is eased, as bunches may be held at the top energy of 10 GeV.

- It can utilize sturdy metallic vacuum chambers, in contrast with an RCS, which must have ceramic chambers with RF shields to limit the eddy currents.
- Single booster and driver rings and transfer lines can be used, saving cost.
- There is low beam power loss during H⁻ injection and easier bunch compression compared with an option that uses a linac, an accumulator and a compressor ring [5].

The NFFAG cell layout is shown in Fig. 4. Three magnet types are used in the basic cell; the bd and BD units are non-linear, vertically focusing, parallel edged, combined function magnets, with bd and BD providing reverse and positive bending, respectively. The F magnet is a non-linear, horizontally focusing, positive bending, combined function magnet, whose edges are parallel to those of bd and BD. Beam loss collimation is a major design issue. The fractional loss in the collimators must be kept below $1 \times 10^{-3}$, with that in the extraction region and elsewhere in the ring both less than $1 \times 10^{-4}$. Halo growth must thus be limited. A layout of the complete proton driver system is shown in Fig. 5.

Figure 6 shows 10 GeV lattice functions, and Table 4 gives orbit data at three energies, 3, 5.9, and 10 GeV. The 3 and 10 GeV orbit separations are largest in the bd unit, reaching 0.33 m. The F unit has a peak orbit field of 1.75 T. Note, however, that the full non-linear magnetic field data were not used in the orbit assessment. Parameters for the adiabatic bunch compression were already given in Table 3. The extraction system comprises a multi-pulse kicker and a septum magnet; it is less demanding than that for the decay rings. Unnormalized beam emittances are similar for protons in the booster and muons in the decay ring, but those for protons in the NFFAG ring are much lower than those for $\mu^{\pm}$ in the 10–20 GeV ring.

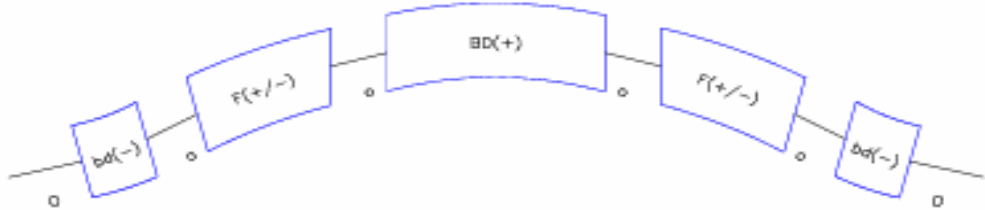

Fig. 4. A single lattice cell of the 50 Hz, 4 MW, 10 GeV, NFFAG proton driver ring.

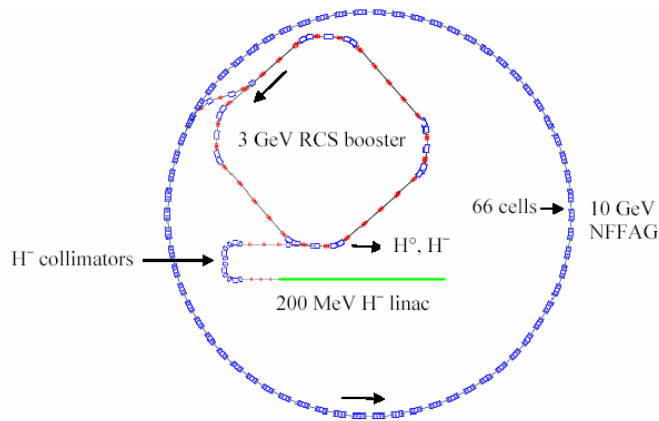

Fig. 5. Schematic layout drawing of the linac, booster and NFFAG of the proton driver.

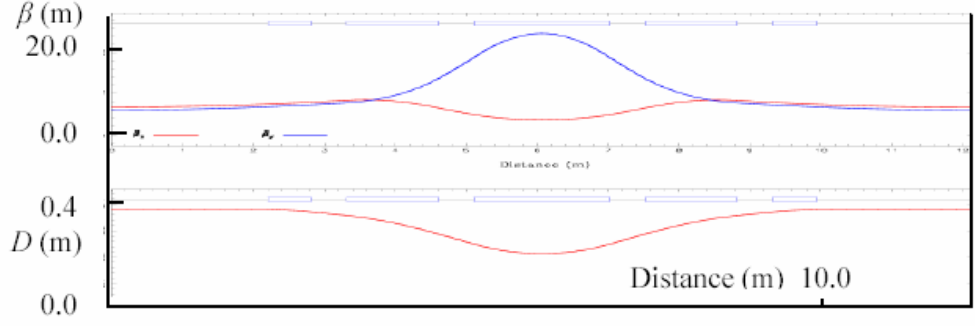

Fig. 6. Small amplitude betatron and dispersion functions for the NFFAG 10 GeV orbit.

Table 4. Orbit parameters at several energies.

| Energy (GeV) | 3.0 | 5.9 | 10.0 |
|---|---|---|---|
| Orbit length (m) | 801.6 | 801.0 | 801.4 |
| Cell length (m) | 12.15 | 12.14 | 12.14 |
| Length of long straight (m) | 4.40 | 4.40 | 4.40 |
| Betatron tunes ($Q_y$, $Q_x$) | 15.231, 20.308 | 15.231, 20.308 | 15.231, 20.308 |
| Gamma transition | −18.9 | 47.7 | 21.9 |
| bd normalized gradient (m$^{-2}$) | 0.059 | 0.029 | 0.005 |
| F normalized gradient (m$^{-2}$) | −0.283 | −0.272 | −0.260 |
| BD normalized gradient (m$^{-2}$) | 0.273 | 0.284 | 0.284 |
| bd bend angle (mrad) | −72.5 | −44.3 | −28.8 |
| F bend angle (mrad) | 8.6 | 37.8 | 62.0 |
| BD bend angle (mrad) | 222.9 | 108.1 | 28.8 |
| bd orbit length (m) | 0.62 | 0.62 | 0.62 |
| F orbit length (m) | 1.285 | 1.285 | 1.290 |
| BD orbit length (m) | 1.924 | 1.921 | 1.920 |
| F to bd orbit length (m) | 0.501 | 0.500 | 0.500 |
| F to BD orbit length (m) | 0.503 | 0.501 | 0.500 |

Twenty four reference orbits are defined for a lattice cell of the 3–10 GeV proton NFFAG. The magnetic field profiles of the F and BD units, together with those of the bd units, are set for zero chromaticity at each reference energy. Gamma transition is 21.86 at 10 GeV, which assists the bunch compression. The non-linear, non-scaling aspects of the ring cause $\gamma_t$ to vary with energy, despite the constant tunes. Gamma transition is imaginary at low energy, real after mid-cycle, and decreases at high energy. A full analysis, not yet done, requires using the full non-linear magnet data, followed by ray tracing in a 6-D simulation program such as ZGOUBI [7].

### *2.5 Linac Option*

The low energy part of a proton accelerator complex always uses a linac. Above a certain kinetic energy, however, conventional setups make use of circular accelerators, which provide a cost advantage because of their efficient use of RF systems. In the case of a multi-MW proton driver, that practice is worth reconsidering because of the need for fast acceleration (requiring fast cycling magnets, ceramic vacuum chambers and an expensive wide-frequency-range RF system providing a lot of voltage), which dramatically increases the cost of the circular accelerator.

Today's linacs are capable of reliably providing tens of MW of beam power. For a Neutrino Factory, however, fixed-energy rings remain necessary to transform the long linac beam pulses into the required number of short bunches.

Linac-based proton drivers are being considered at FNAL [8] and at CERN [9]. The CERN proposal will be used as a typical example. The linac itself is a modified version of the 3.5 GeV Superconducting Proton Linac (SPL), whose Conceptual Design Report has recently been published [10]. Longer by 105 meters and equipped with 14 more 4 MW klystrons (for a total of 58), it accelerates protons up to 5 GeV in an overall length of only 534 m. Its structure is sketched in Fig. 7 and its main characteristics are shown in Table 5.

Two fixed-energy rings of approximately 300 m circumference are necessary to give the proton beam the required time structure for a Neutrino Factory. In the first one, the 400 μs linac beam pulse is accumulated using charge-exchange injection. Once accumulation is finished, bunches are transferred to a compressor ring where they are rotated in the longitudinal phase plane and ejected to the target when their length is minimum.

The mode of operation of the accumulator and compressor rings is illustrated in Fig. 8. The time structure of the chopped linac beam is chosen such that the beam circulating in the accumulator forms 5 bunches. The accumulator is designed to be quasi-isochronous ($\gamma_t^2 \sim 50$, making the phase-slip factor $\eta \sim 0.02$), so no RF is necessary to preserve the bunches during the full accumulation and compression process. At the end of accumulation, bunches are successively sent to the compressor ring every 12-1/5 turns. The ratio of the circumferences is selected such that the bunches arrive spaced by one-third of a turn in the compressor. Inside the compressor, bunches rotate in the longitudinal phase plane under the action of a 4 MV, $h = 3$ RF system. After

Table 5. Characteristics of the 5 GeV version of the SPL.

| | |
|---|---|
| Ion species | $H^-$ |
| Kinetic energy (GeV) | 5 |
| Beam power (MW) | 4 |
| Repetition rate (Hz) | 50 |
| Mean current during the pulse (mA) | 40 |
| Pulse duration (ms) | 0.4 |
| Bunch frequency (MHz) | 352.2 |
| Linac length (m) | 534 |

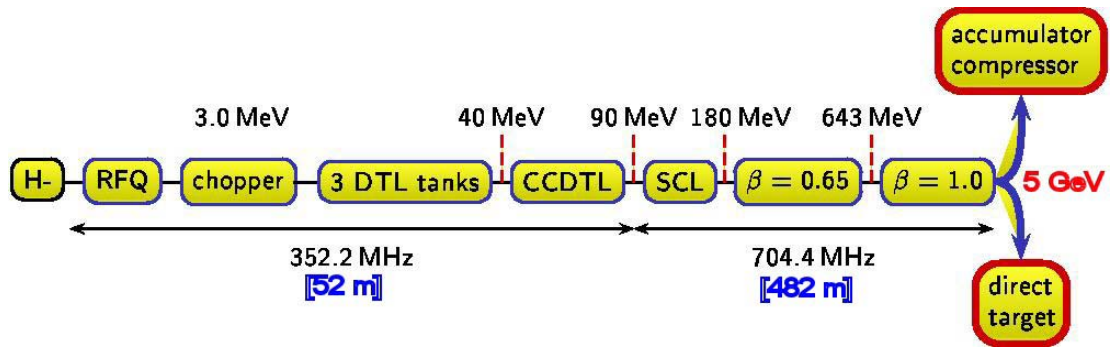

Fig. 7. SPL block diagram.

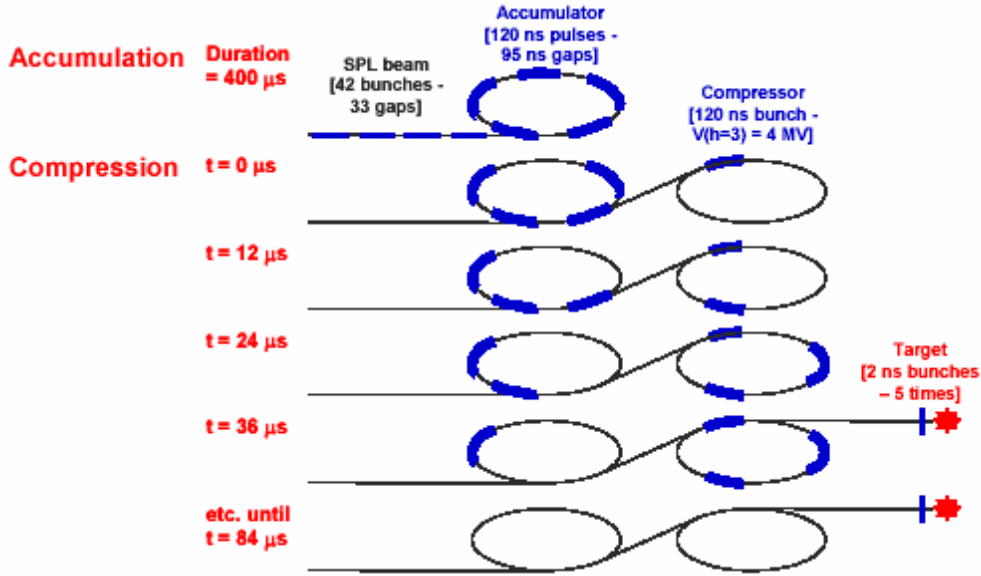

Fig. 8. Accumulation and compression scheme for linac-based proton driver.

36 turns, the first bunch has a minimal length of approximately 2 ns, whereupon it can be ejected to the target. The following bunches are then successively ejected every 12-1/3 turns, for a total burst length of ~50 $\mu$s.

### 2.5.1 Superbeams

In the case of Superbeams, the neutrino beam comes from the decay of pions and the muons themselves are not used. The infrastructure beyond the target is thus limited to a focusing system and a decay tunnel. Therefore, every laboratory equipped with high energy proton accelerators has been the subject of Superbeam proposals, and very different energies for the primary proton beam have been assumed, ranging from 3.5 to 400 GeV. The neutrino flux being directly proportional to beam power, the corresponding accelerators have to deliver a very high flux, sometimes well beyond what they were initially designed for. A representative list is given in Table 6.

Compared with a Neutrino Factory, the only requirements on the time structure of the proton beam come from the need to sufficiently reject background in the remote experiment. For that purpose, the duty factor of the proton beam has to be lower than $5 \times 10^{-3}$ [11].

### 2.5.2 Beta Beams

The first study of a Beta-Beam [12] facility within the EURISOL design study is not yet completed, but it has already become apparent that the main challenges of such a facility include radioactive ion production, beam bunching, and collimation and magnet protection in the acceleration and decay rings. In Europe, a design study for a neutron facility has begun, and Beta Beam studies will be carried out under its auspices. The Beta Beam work will focus on:

- an in-depth study of a production ring for high intensities of radioactive ions [13]
- continued tests of the EURISOL 60-GHz ECR source for bunching

Table 6. Proposed Superbeams.

|  | Proton beam energy (GeV) | Protons per pulse | Repetition period (s) | Beam power (MW) |
|---|---|---|---|---|
| CNGS+ [16] | 400 | $4.8–14 \times 10^{13}$ | 6 | 0.3–1.2 |
| FNAL [17] | 120 | $9.5–15 \times 10^{13}$ | 1.5 | 1.1–2 |
| JPARC [18] | 50 | $33 \times 10^{13}$ | 3.64–1.6 | 0.6–1.5 |
| BNL [19] | 28 | $9–25 \times 10^{13}$ | 0.4–0.1 | 1–4 |
| FREJUS [20] | 3.5 | $14.3 \times 10^{13}$ | 0.02 | 4 |

- machine studies and particle-matter interaction simulations of the accelerator and decay ring[3] collimation and magnet protection systems

The objectives are to propose a solution for the shortfall of $^{18}$Ne in the EURISOL design study of a beam for a $\gamma = 100$ Beta-Beam facility on a 130 km baseline (e.g., CERN to Frejus) or for a $\gamma = 350$ facility with a long baseline [14], and to study the use of high-$Q$ ($^{8}$Li and $^{8}$B) isotopes for a $\gamma = 100$ facility with a long baseline (e.g., CERN to CNGS)[15].

In general, the proton driver demands for a Beta Beam facility are modest compared with those for a Neutrino Factory or Superbeam facility. The Beta Beam production technique in most cases limits the proton beam power to the level of tens of kW, as opposed to the MW-level proton beams needed for a Neutrino Factory or Superbeam. Moreover, the process of capture and reionization of the radioactive species completely decouples the beam properties from that of the proton driver. Although the Beta Beam design is a challenging one, from an accelerator design perspective it has little, if any, commonality with those for a Neutrino Factory or Superbeam.

## 3. Target Issues

### 3.1 Beam Energy Choice

To determine the kinetic energy of the proton beam that is most efficient for the production of soft pions, we process the produced pions through the entire front end of the Neutrino Factory using the Study 2a [21] configuration. As a figure of merit, we select surviving muons that are fully contained within the capture transverse acceptance (30 $\pi$ mm-rad) and the longitudinal acceptance (150 mm) assumed for the subsequent accelerating section. The particle production model used was MARS V14 [22] and the propagation of the particles though the Neutrino Factory front end was done utilizing the ICOOL code [23]. The efficiency of the muon capture was computed by evaluating the number of collected muons at the end of the Neutrino Factory front end and normalizing the results to the power of the proton beam. Results utilizing a mercury-jet target are shown in Fig. 9. Target parameters such as radius, tilt angle, and longitudinal placement were previously optimized in Study 2a [21].

---

[3] Study of the decay ring will include measurement of relevant cross sections for such a ring and tests of its radioactive ion collection system.

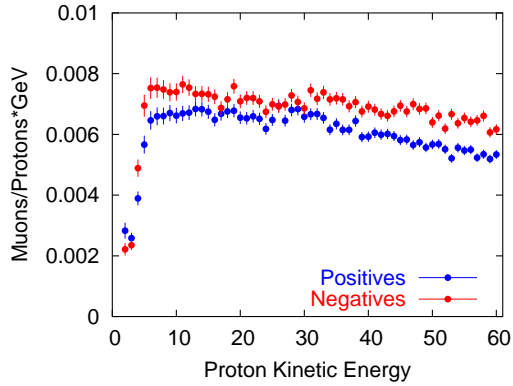

Fig. 9. Calculated production efficiency of positive and negative muons at the end of the Study 2a cooling channel, per proton and per GeV of proton beam energy, for a mercury-jet target. Although the curves are rather flat, an optimum energy, roughly 10 GeV, is discernible. Below about 5 GeV, the calculations show an abrupt fall-off in production. Above 10 GeV the fall-off is small but, from the muon production perspective, there is no benefit to increasing the beam energy beyond 10 GeV.

## *3.2 Choice of Target Material*

We also investigated other candidate target types. Figure 10 shows an efficiency plot for a carbon target. Here, the optimal proton kinetic energy is centered around 5 GeV, somewhat lower than the case for mercury. As can be seen from comparing the two figures, the high-*Z* material shows the higher efficiency for soft-pion production, which will lead to the greatest number of captured muons. In evaluating the most efficient kinetic energy region for a mercury target, we find that 6–38 GeV protons give a sum of positive and negative pions within 10% of the maximum efficiency at 10 GeV.

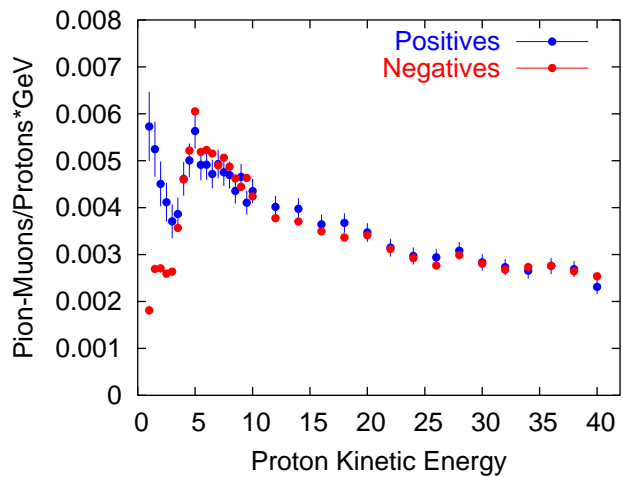

Fig. 10. Calculated production efficiency for a carbon target. The yield per proton and per GeV is lower than for a mercury target and peaks at a lower energy.

### *3.3 Proton Beam Structure*

### 3.3.1 Repetition Rate

For a given proton driver power, an increased repetition rate will lower the stress on the target (especially for a solid target) since the intensity per pulse is decreased. For the same pulse intensity and increased repetition rate, the proton driver power increases with a concomitant increase in stress on the target.[4] The primary downside of a higher repetition rate is the increased average power consumption of the RF systems. In Study 2 [24], the average power required for these systems was 44 MW for a 15 Hz average repetition rate. This portion of the machine's power consumption will be proportional to the repetition rate.

Higher repetition rates reduce the amount of current per bunch train, which in turn reduces the beam loading in the RF cavities. Furthermore, some schemes for the storage ring require (superconducting) RF cavities to keep the beam bunched, and higher currents would require more RF power (and possibly more cavities) to compensate for beam loading there.

### 3.3.2 Pulse Length, Intensity, and Structure

The pulse intensity, combined with the beam spot size, controls the quasi-static conditions of pressure and temperature generated in the target. Energy densities of up to 400 J/g, corresponding to ~ $24 \times 10^{12}$ protons per pulse and $\sigma_r = 1$ mm, may be tolerated by some high performance solid materials. The pulse length controls the ensuing dynamic stresses and can play a significant role in determining whether a solid target survives the induced shock. Solid targets favor longer pulses because of the ability to relax during deposition. On the other hand, a liquid-jet target performs best at very short pulse lengths (a few ns), as the onset of jet destruction occurs much later. At the same intensity, a pulse having a uniform distribution over the same area as a Gaussian pulse (i.e., a $3\sigma$ spot) will reduce the stress and temperature in the target by approximately a factor of three.

### 3.3.3 Bunch Length

The proton bunch length has a strong influence on the usable muon intensity. The accepted muon density at the end of the cooling channel falls off with increasing proton driver bunch length on the target. This behavior can be partially understood by a simple theory that models the longitudinal dynamics of the muon beam through the RF components of the front end. Longer proton bunches produce initial longitudinal phase space areas that exceed the longitudinal acceptance of the front end.

### *3.4 Solid Target Considerations*

There are a number of problems to be addressed with a solid target:

- the effect of thermal shock produced by short pulses of a high intensity proton beam
- the ability to cool the target at proton beam power above ~1 MW
- extreme radiation damage in the target
- eddy current effects associated with moving the target material through a 20-T solenoid

---

[4]That is, the thermal load of each pulse on the target must be removed by the heat sink in a shorter time and the repetition rate will be limited by the ability to remove the dynamic stresses entirely between pulses.

### 3.4.1 Solid Target Implementation Scenario

Figure 11 shows schematically one scenario proposed [25] for solid targets. High-$Z$ target bars, 2–3 cm in diameter and 20–25 cm long, are moved through the beam at a velocity of 5 m/s. The bars, which operate at ~1800 K and radiate the power dissipated in them to water-cooled walls, are connected to a pair of chains that propel and guide them through the solenoids. If there are ~500 bars on the chains, each target bar passes through the beam only one million times in one year of operation. Moving the bars transversely across the beam avoids reabsorbing the pions in the "downstream" target material. Of course, there are many mechanical implementation issues associated with such a scheme that have not yet been examined. Detailed studies are still needed before such an approach could be considered feasible.

Tantalum was originally chosen for the target material since it was observed to behave well under proton irradiation at ISIS, suffering negligible damage up to 12 dpa. ISIS now uses tungsten and this appears to behave equally well.

### 3.4.2 Thermal Shock Studies

To evaluate the feasibility of a solid target, experiments [26] on thermal shock have been carried out by passing a high current pulse through thin tantalum and tungsten wires at high temperatures. Figure 12 shows the setup and Fig. 13 a photograph of the wire assembly mounted on the insulating vacuum feedthrough.

Calculations using LSDYNA predict the stress expected in the target bars with 4 MW of incident beam power. The proton beam was assumed to consist of macro-pulses at 50 Hz repetition rate, each of which contains one or more microbunches. The stress is reduced if there are more microbunches and, of course, by a larger beam diameter[5]. As shown in Fig. 14, spacing the bunches appropriately reduces the stress. The current design concept utilizes 3 bunches, 2 ns long, separated by ~10 µs. This arrangement, with comparatively large spacing, also helps ease some design challenges for both the proton and the muon accelerators.

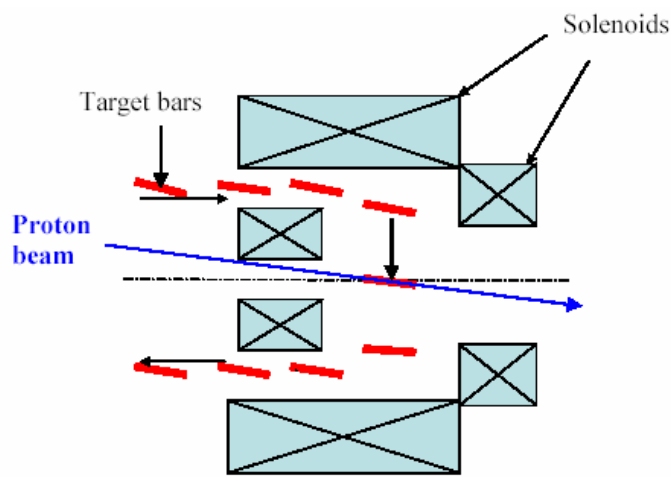

Fig. 11. Schematic diagram of a solid target scenario.

---

[5] The beam is assumed to have a parabolic distribution with diameter equal to that of the target.

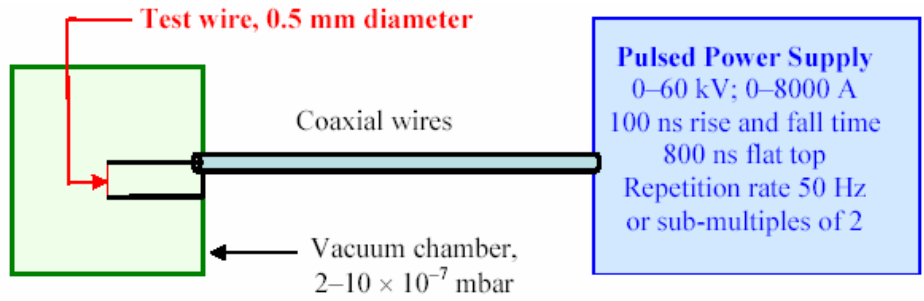

Fig. 12. Schematic diagram of the test wire and power supply.

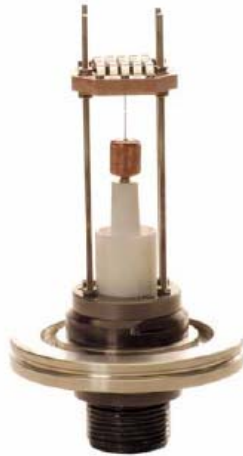

Fig. 13. Wire test assembly.

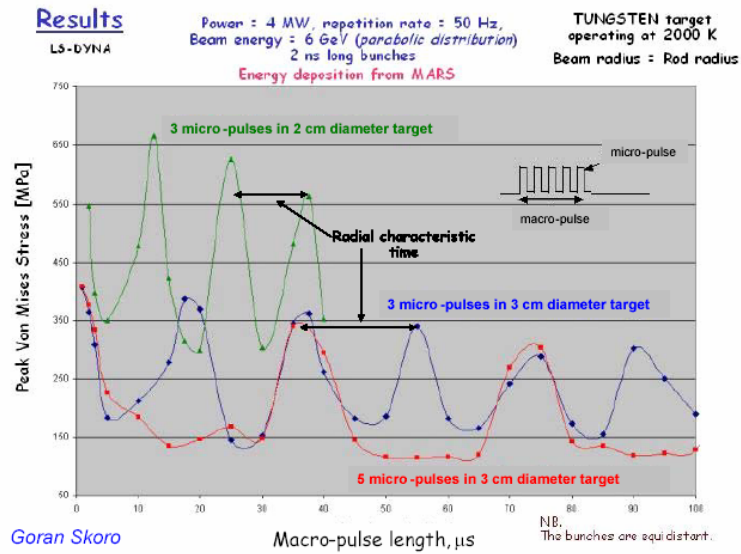

Fig. 14. Thermal stress as a function of target diameter, number and spacing of proton bunches in a macro-pulse.

To test the concept, an electrical current was employed that generates the same peak stress in the wire, including both Lorentz and thermal forces (see Fig. 15). Initial experiments [26] showed that tantalum was too weak to withstand more than a few hundred thousand pulses at temperatures over 1000 K. However, tungsten is much stronger at high temperatures and has withstood over 26 million pulses at a stress equivalent to >4 MW into a 2 cm diameter target at 1900 K. These experiments are continuing and will include some high strength alloys of tungsten (see Fig. 16) and graphite. Radial and longitudinal surface acceleration of the wire will be measured with a VISAR (Visual Image Stabilization and Registration) system so that the equations of state of the wire at high temperature under shock conditions can be assessed using the models in LSDYNA.

### 3.4.3 Conclusions from the Wire Tests

Thermal shock from one or a few beam pulses does not appear to be a major problem for a suitable solid-target material, but fatigue and creep likely are. These are difficult parameters to assess in terms of lifetime, but the present wire tests indicate that over ten million pulses (>10 years of life) can be expected. The effects of beam misalignment and implications of shape changes and thermal distortions on the delivery system have yet to be assessed.

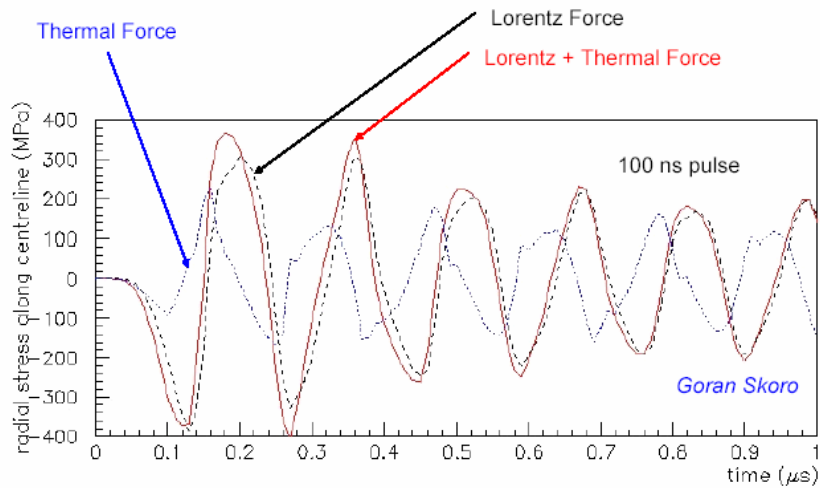

Fig. 15. Typical radial stress in the wire from thermal and Lorentz forces.

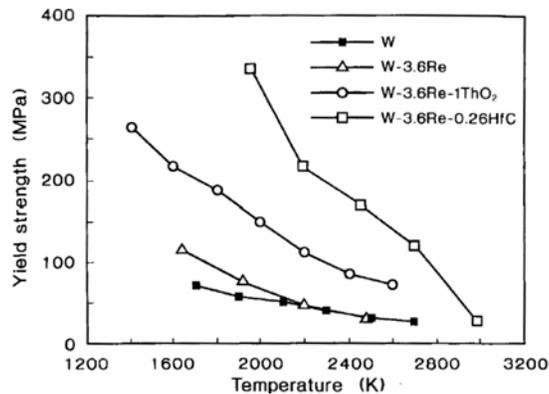

Fig. 16. Yield strength of tungsten and some its high-strength alloys.

### *3.5 Baseline Configuration*

After evaluating the available results, we adopted the Study 2a [21] configuration, with a liquid-Hg jet, as our baseline configuration. The system is illustrated in Fig. 17. It comprises a 20-T hybrid solenoid (with a superconducting outer coil and a resistive inner coil), followed by a series of superconducting solenoids that taper the field adiabatically down to 1.75 T. The proton beam enters the solenoid at a 67 mrad downward angle with respect to the solenoid axis, and the Hg jet has a 100 mrad angle. The 33 mrad angle between the beam and the Hg jet was shown in earlier work to optimize pion production by reducing reabsorption of the newly created pions.

The Hg jet is a closed-loop flowing system, driven by a remote pump. The beam dump is a pool of Hg that forms part of the overall target loop. The Hg from the dump is circulated through a heat exchanger to remove the power deposited by the beam. One advantage of the liquid-Hg system is that the target material can be purified by distillation, removing many of the radionuclides produced by the proton beam.

## 4. Front End

The parts of the Neutrino Factory between the target and the beginning of the acceleration system are designated collectively as the *front end* [27]. There are two main requirements on the operation of the front end. First, it has to collect the pions created in the target and form a beam from their daughter muons as efficiently as possible. Second, it has to manipulate the transverse and longitudinal phase space of the muon beam so that it matches the accelerator acceptance as efficiently as possible. The Neutrino Factory front end described here is made up of the following subsystems:

- π/µ collection
- π decay region
- bunching
- phase rotation
- ionization cooling

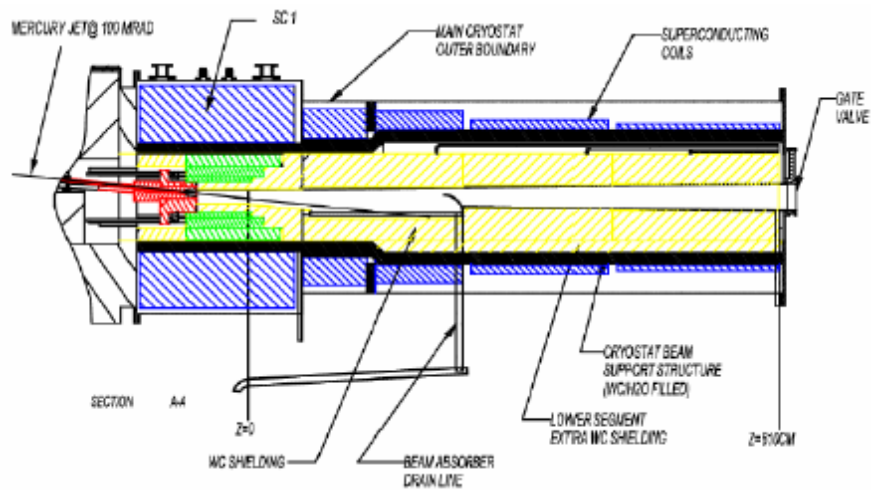

Fig. 17. Baseline ISS target system. The solenoidal field at the target position is 20 T. The downstream coils begin the process of adiabatically tapering the field down to 1.75 T for further transport.

The initial transverse phase space of the muon beam is determined mainly by the magnetic field strength in the channel, which provides the required focusing, and the radial aperture of the beam pipe. The longitudinal phase space can be modified by allowing the beam to travel a long distance in an empty magnetic lattice. This permits a correlation to develop between the temporal position and the energy of the particles in the bunch. Electric fields in RF cavities are then used to rotate the longitudinal phase space. This produces a longer particle bunch with a reduced energy spread. To ensure efficient acceleration, it is necessary to bunch the beam to match the frequency of downstream RF cavities. Finally, it is also necessary to decrease the transverse emittance of the collected beam by means of an ionization cooling channel in order to optimize Neutrino Factory intensity.

Much of the tracking done in our ISS front end simulations was done using the ICOOL code [23, 28]. This program does beam tracking in accelerator coordinates. The fields come from built-in magnet and RF cavity models, from field maps, or from files of Fourier coefficients. The code accurately models the decays of particles and their interactions in matter, including energy loss, energy straggling, and multiple Coulomb scattering. A similar code MUON1 [29] has been used for front end simulations at Rutherford Appleton Laboratory.

### *4.1 Comparison of Front-end Systems*

To understand the advantages and disadvantages of various cooling approaches, we compared the cooling channels proposed in the three published Neutrino Factory feasibility studies [21, 27, 30].

The single most significant difference in the various approaches is the choice of RF frequency:

- Japanese FFAG study [31], 5 MHz
- CERN linear channel studies [32–38], 88 MHz
- U.S. linear channel studies [21, 24, 39–41], 201 MHz

Another key choice involves the method of longitudinal capture, using either a single bunch in one RF bucket or a train of many bunches. This choice depends not only on the RF frequency but on the bunch structure of the proton driver. Because ionization cooling was included in most, but not all, designs, comparisons were made both with and without this feature. To permit valid performance comparisons, the ISS baseline decay ring configuration—a racetrack ring—was used for all cases. Although much of our analysis is based on the published studies, we have also considered some variants not included by the original designers.

The analysis was based on the muon capture rate per initial pion in the decay channel, and this final efficiency is given in Table 7. This approach avoids the complication of pion production uncertainties, and particularly their energy dependence. It is, however, useful to relate these efficiencies to a number of muons per year with one fixed assumption of the pion production per proton per GeV. To do this, the following assumptions were made:

1. the number of captured pions per 24 GeV proton in Study 2a, 0.94, was used; this can be expressed in terms of pions per proton, per GeV of proton energy, as 39%

Table 7. Summary of efficiencies for different cases, including an estimate of useful muon decays per year assuming the same pion production estimate for 10 GeV protons. Parenthesized values are unpublished estimates or calculations made for this comparison study. The + sign indicates the inclusion of a system to separate and separately phase rotate each sign.

| $f_{RF}$ (MHz) | Ref. | Cool | $A_\perp$ (mm-rad) | Phase rotation | $\eta_\perp$ (%) | $\eta_\parallel$ (%) | $\eta_{accel}$ (%) | $n_\pm$ | $\eta_{all}$ (%) | μ/year ($10^{21}$) |
|---|---|---|---|---|---|---|---|---|---|---|
| 5 | [31] | No | 30 | No | (18) | (39) | 50 | 1 | 3.5 | 0.11 |
| 5 | [31] | No | 30 | Yes | (18) | (60) | 50 | 1 | 5.4 | (0.17) |
| 5 | [31]+ | No | 30 | Yes | (18) | (60) | 50 | 2 | 11 | (0.34) |
| 44-88 | [32–35] | Yes | 15 | Yes | (50) | (15) | 65 | 1 | 4.9 | 0.16 |
| 44-88 | [32–35] | Yes | 15 | Neuffer | (50) | (48) | 65 | 2 | 31 | (1.0) |
| 44-88 | [32–35] | No | 30 | Neuffer | (20) | (48) | 65 | 2 | 13 | (0.41) |
| 201 | [40] | Yes | 15 | Multi | 31 | 56 | 81 | 1 | 14 | 0.45 |
| 201 | [40]+ | Yes | 15 | Multi+ | 31 | 56 | 81 | 2 | 28 | (0.9) |
| 201 | [40] | No | 30 | Multi | 24 | 56 | 81 | 1 | 11 | 0.35 |
| 201 | [21, 41] | Yes | 30 | Neuffer | 42 | 48 | 81 | 2 | 33 | 1.06 |
| 201 | [21, 41] | No | 30 | Neuffer | 24 | 48 | 81 | 2 | 19 | 0.61 |

2. to correct for the MARS predicted improvement in performance at 10 GeV compared with the Study 2a choice of 24 GeV, the pion yield was increased by 10%
3. for the racetrack ring geometry, 38% of decays in the ring are assumed to take place in the production straight section of the decay ring
4. an average proton beam power of 4 MW was taken
5. a "Snowmass year" of $10^7$ s was assumed

As can be seen in Table 7, the basic U.S. Study 2a scheme using 201 MHz is the only proposal that appears to meet the intensity requirement without modification. It does somewhat better than an 88 MHz scheme employing multi-bunch phase rotation on account of its smaller decay losses during acceleration due to greater accelerating gradients. This scheme has less transverse acceptance (42%) than the CERN 88 MHz case (50%) because it has less cooling.[6] Details of how the values in Table 7 were estimated are covered in the subsections below.

### 4.1.1 Analysis Method

The overall production and capture efficiency could be defined as the number of captured and accelerated muons per unit of proton energy. Unfortunately, such a definition is dependent on the assumed pion production rates, which are not consistent with one another (see Table 8). The production also depends on the proton energy used—a parameter that is not relevant when comparing capture, cooling, and acceleration methods.

---

[6]This was a deliberate choice to reduce cost. If more cooling, using a tapered channel and liquid hydrogen (as in Study 2 and the CERN proposal) were used, somewhat higher performance could be achieved, though at significant incremental cost.

Table 8. Pion and muon production estimates in different cases.

| Case | Reference | Program | $E_p$ (GeV) | $p$ Range (GeV/c) | $\mu/\pi$ | $\mu/\pi$ per GeV |
|---|---|---|---|---|---|---|
| FFAG Solenoid | [31] | MARS | 50 | | 1.2 | 0.024 |
| FFAG Solenoid | [31][a] | MARS | 50 | 0–1000 | 2.0 | 0.040 |
| CERN Solenoid | [32] | FLUKA | 2.2 | 50–800 | 0.18 | 0.082 |
| CERN Solenoid | [34] | MARS | 2.2 | 50–800 | | 0.010 |
| CERN 300 kA Horn | [34] | MARS | 2.2 | 50–800 | | 0.014 |
| CERN 400 kA Horn | [34] | MARS | 2.2 | 50–800 | | 0.017 |
| CERN Solenoid | [34] | MARS | 16 | 50–800 | | 0.025 |
| Study 2a | [21, 41] | MARS | 25 | 0–1000 | 0.8 | 0.033 |

[a] See Fig. 2.6 in Ref. [32].

To avoid these ambiguities, we define here an efficiency, $\eta$, as the number of final muons divided by the number of pions and muons captured by the initial solenoid and transported at least 1 m down the decay channel. We then further express this efficiency as an approximate product of three sub-efficiencies:

$$\eta = \frac{\text{Final muons}}{\text{Pions in decay cahnnel}} = \eta_{\parallel} \eta_{\perp} \eta_{\text{accel}} \quad (1)$$

For the purposes of this study, we will include decay losses in the phase rotation scheme in the efficiency $\eta_{\parallel}$, the decay losses in the cooling in $\eta_{\perp}$, and the decay losses during acceleration in $\eta_{\text{accel}}$. In addition, we will define a 'front-end' efficiency as

$$\eta_{\text{front}} = \eta_{\parallel} \eta_{\perp}. \quad (2)$$

Another key difference between cases is the number of muon signs captured. In conventional phase rotation or matching, only one sign is matched (see Fig. 18a), but in Neuffer's bunched beam phase rotation (see Figs. 18b and 18c) [1], both signs are captured with good efficiency. As most neutrino experiments need to study both neutrino and anti-neutrino reactions, the capture and decay of both signs simultaneously effectively doubles the overall efficiency.

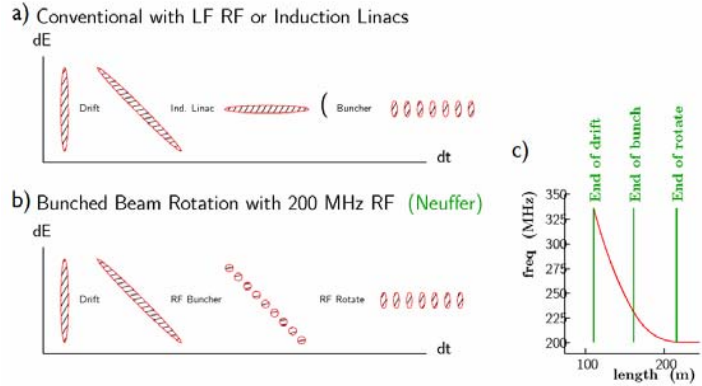

Fig. 18. Phase rotation methods: a) conventional; b) Neuffer bunched beam phase rotation. c) Shows frequency *vs.* distance for the case of Study 2a, a version of b).

In what follows, some of these efficiencies are derived directly from the published studies, while others, to be shown in parentheses, have been estimated from separate calculations or simulations.

### 4.1.2 Production and Capture

Before proceeding with the performance comparison, it is instructive to look at the assumed pion production in the different studies. This is reasonably straightforward, as most studies included the use of a 20 T, 7.5 cm radius solenoid and adiabatic matching to a solenoidal decay channel (see Figs. 17 and 19).

For particles starting on the axis, the maximum transverse momentum of tracks captured in a solenoid of field $B$ and radius $r$ is

$$p_\perp \text{ (GeV/c)} = \frac{cBr}{2} \approx 0.15 B\text{ (T)}\, r\text{ (m)}. \tag{3}$$

The normalized acceptance is then

$$A_n (\pi \text{ m-rad}) = \frac{Br^2 c}{2 m_\mu} = 1.43 B\text{ (T)}\, r^2\text{ (m}^2\text{)}, \tag{4}$$

which, for 20 T and a radius of 7.5 cm, is 160 $\pi$ mm-rad. The acceptance of the decay channel into which this initial acceptance must be adiabatically matched is different for each case (see Table 9), but all are greater than this initial acceptance and cause no loss of the originally captured particles. We can thus use the published numbers of pions and muons at least 1 m from the target to compare the assumed production capability. These published production estimates are summarized above in Table 8. Although the momentum ranges used differ somewhat, very few particles lie outside those ranges, so such differences are not very significant.

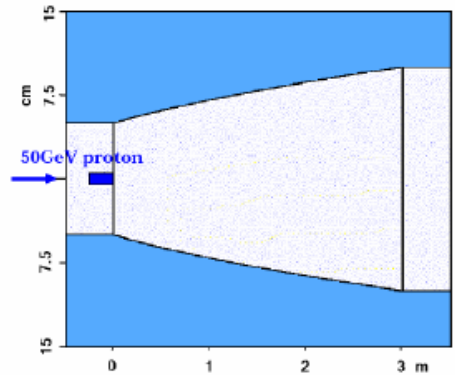

Fig. 19. Capture system from Ref. [31], comprising a 20 T solenoid followed by a rapid taper to 5 T.

Table 9. Acceptances of decay channels.

| Case | Mean pion momentum (MeV/c) | Decay field (T) | Decay radius (cm) | Acceptance ($\pi$ mm-rad) |
|---|---|---|---|---|
| FFAG | 300 | 5.0 | 16 | 180 |
| CERN | 286 | 1.8 | 30 | 250 |
| FS2 | 220 | 1.25 | 30 | 170 |
| FS2a | 220 | 1.75 | 30 | 240 |

It is apparent that there are inconsistencies between the various simulated production models even at the same proton energies. Under these circumstances, comparing the final claimed muon rates per proton energy would be substantially distorted by the differing pion production assumptions. As the object of this exercise is to compare different methods of capture, cooling, and acceleration, we have chosen to compare muon fluxes normalized to the fluxes of pions and muons captured in the 20 T solenoids and transported down the decay channels.

Table 8 also includes two cases using magnetic horn capture (see Fig. 20) in place of solenoid capture. In both cases, using the same production assumptions, the capture is less than with the solenoid. As there was no attempt made in these horn-based studies to match the resulting particles into the decay channel, there could be additional losses introduced by the matching.

### 4.1.3 RF Choices and Accelerating Gradients

The FFAG study [31] considered very large 5 MHz vacuum RF cavities (Fig. 21) and also ferrite loaded systems (Fig. 22). In the former case, the achievable accelerating gradient is limited by surface breakdown, and in the latter case by ferrite performance. Gradients of 1 MV/m were assumed in Ref. [31], but recent discussions have suggested 0.75 MV/m as more realistic. Although neither gradient has yet been demonstrated in this configuration, induction linacs, which have similarities to the proposed ferrite-loaded cavities, have achieved over 1 MV/m.

A system was studied at CERN [32–35] using 44 MHz at the beginning of the capture, phase rotation and cooling channel, followed by 88 MHz (see Fig. 23) after initial cooling. A later study [36, 37] considered 88 MHz for the whole front end, achieving the same performance. Full details of the latter study are not available. Average accelerating gradients of 2.5 MV/m at 44 MHz and 4 MV/m at 88 MHz were assumed.

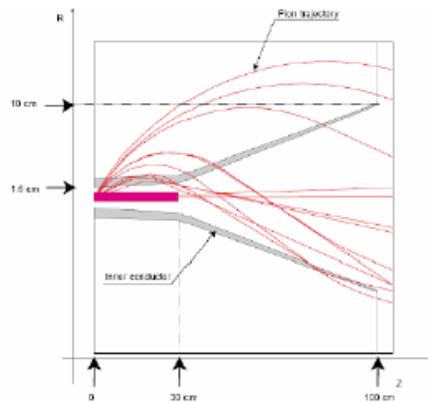

Fig. 20. CERN horn capture option.

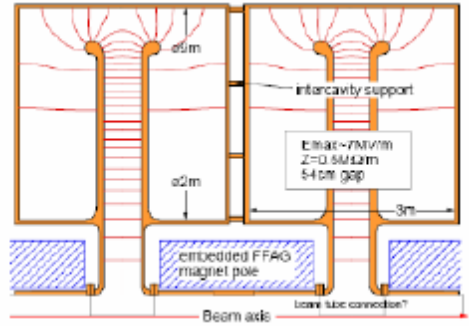

Fig. 21. 5 MHz vacuum cavity for FFAG scheme.

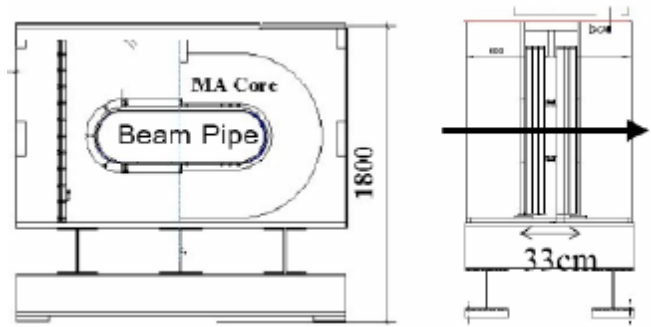

Fig. 22. 5 MHz ferrite-loaded cavity for FFAG scheme.

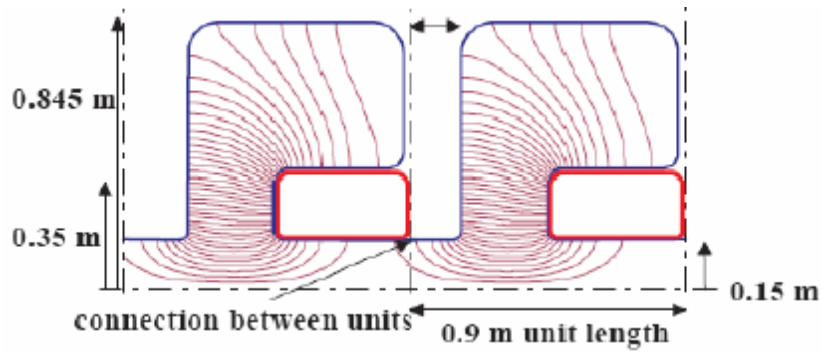

Fig. 23. 88 MHz vacuum cavity with integrated solenoids.

A number of recent studies [21, 24, 39–41] have used 201 MHz with gradients up to 16 MV/m. A copper cavity (see Fig. 24) has achieved this gradient without difficulty in the absence of external magnetic fields, but reaching it when immersed in a solenoidal field remains to be demonstrated. A superconducting 201 MHz cavity has achieved 11 MV/m at Cornell, but was limited from operating at higher gradients by unexpectedly high losses. It is hoped that further development will solve this problem and allow the gradients specified for the acceleration system of 17 MV/m.

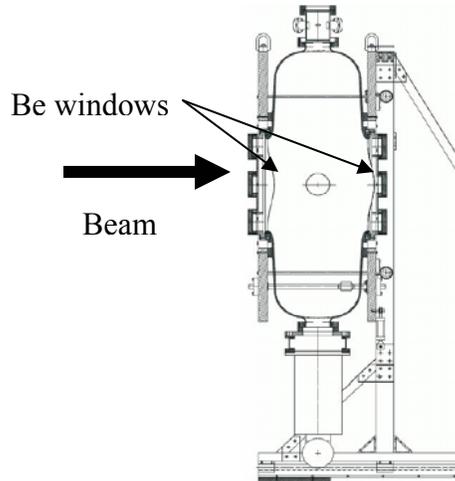

Fig. 24. 201 MHz test cavity, showing mounting arrangement. The ports on the entrance and exit faces are for diagnostic purposes. The curved surfaces represent thin Be windows that electrically close the irises.

Parameters for each case examined are summarized in Table 10 and plotted in Fig. 25. We see from the plot that the gradients assumed below 201 MHz are low compared with a simple square-root-of-frequency extrapolation that fits at high frequencies. The reason is that, for vacuum cavities below 200 MHz, a simple pillbox shape would be impractically large. Reducing cavity size requires shapes having enhanced surface fields. Breakdown in the gap limits the practical gradient. For ferrite-loaded cavities, gradients are limited by losses in the ferrite.

Table 10. Parameters of assumed RF systems.

| $f_{RF}$ (MHz) | Type | Reference | Gradient (MV/m) | Avg. gradient (MV/m) |
|---|---|---|---|---|
| 5 | Ferrite | [31] | 0.5–2.0 | 0.75–1.0 |
| 5 | Vacuum | [31] | 1.0 | 1.0 |
| 44 | Vacuum | [36] |  | 2.0 |
| 88 | Vacuum | [36] |  | 4.0 |
| 201 | Vacuum | [24, 40] | 16 | 12 |
| 201 | SC Vacuum | [24, 40] | 17 | 10 |

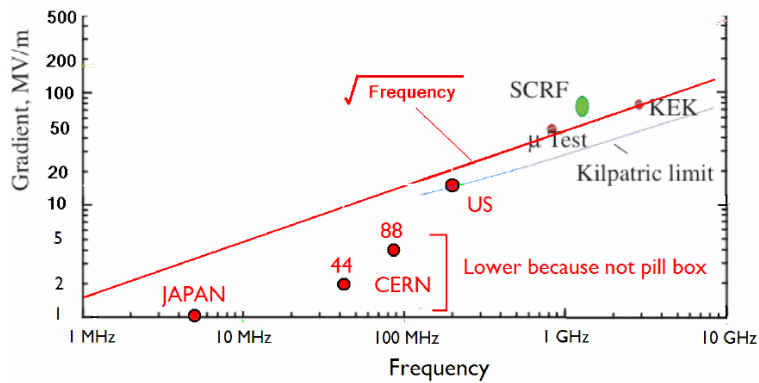

Fig. 25. Gradients vs. frequency.

### 4.1.4 Longitudinal Phase Space Considerations

The challenge is to match the initial muon longitudinal phase space into one or more RF buckets for subsequent cooling and/or acceleration. It is instructive to estimate the effective longitudinal phase area of the initially produced muons, to compare this with the bucket areas of the assumed RF into which they should be matched. We will use longitudinal normalized acceptances defined using the same units as normalized transverse emittances. The normalized longitudinal acceptance for $\beta\gamma \approx 2$, $\Delta E/E = \pm 100\%$, and a time spread of 3 ns is then:

$$A_\parallel = \beta\gamma \frac{\Delta E}{E} c\Delta t \approx 2 \times 1 \times 2 \times 1 = 4 \ (\pi \text{ m-rad}) \tag{5}$$

This is equivalent to 1.3 eV-s, and is a very large acceptance. The bucket area into which we wish to match this acceptance depends on the frequency, gradient, and RF phase angles. The number of buckets into which it can be matched depends on the system of "phase rotation" assumed, and the proton bunch structure. In the 201 MHz case, the proton bunches were well separated, and the muons generated were matched into a long train of about 50 bunches. In the 5 MHz and 88 MHz cases, the published studies matched into only a single RF bucket. Note, however, that in the 88 MHz case, if the proton bunch structure were modified, matching into multiple bunches would also be possible.

Using the RF parameters discussed above, we give the approximate total acceptances for each case in Table 11. We see that, for the 5 MHz case, the bucket area is significantly larger than the beam area so very good acceptance is expected. The same is true of the other cases *as long as the capture is into multiple bunches*. As is obvious from Table 11, only poor acceptance would be possible at the higher frequencies if the matching were into a single RF bucket.

### 4.1.5 Phase Rotation Schemes

Two fundamentally different schemes have been discussed. In the "conventional" approach (Fig. 18a), the initial distribution of particles is allowed to drift and develop a momentum-time correlation. A time-varying voltage, from RF or an induction linac, then decelerates the early, high-momentum particles and accelerates the late, low-momentum particles. The resulting rotated distribution can be captured in a single RF bucket (as in [32, 33]), or bunched to fill a train of bunches (as in [39] or [24, 40]).

In the alternative bunched-beam phase rotation or "Neuffer" method [1], used in [21, 41], the particles are bunched as they drift, prior to phase rotation (see Fig. 18b). To accomplish this bunching, the RF frequency is varied as a function of distance down the channel (see Fig. 18c), such that the RF phase seen by the bunch centers remains constant despite their differing forward velocities. Phase rotation in this case is generated by a judicious shifting of the RF phases of the bunching RF, again such as to decelerate the high-energy early arrivals, and accelerate the low-energy late arrivals. There are two advantages to this scheme:

1. it does not require expensive low frequency RF or induction linacs; and
2. it automatically captures both muon signs, placing them in interleaved buckets as needed for subsequent cooling or acceleration.

Table 11. Longitudinal bunch acceptance in different cases.

| $f_{RF}$ (MHz) | Multi-bunch | No. of bunches | Bucket area ($\pi$ m-rad) | Bucket/Beam Area | Notes |
|---|---|---|---|---|---|
| 5 | No | 1 | 13 | 3.2 | Very good |
| 88 | No | 1 | 0.3 | 0.08 | Bad |
| 88 | Yes | 25 | $0.3 \times 25 = 7.5$ | 1.8 | Good |
| 201 | No | 1 | 0.15 | 0.04 | Very bad |
| 201 | Yes | 50 | $0.15 \times 50 = 7.5$ | 1.8 | Good |

Note that, in either scheme, if the rotation is into multiple buckets, the next proton bunch must be far enough away in time to allow resetting the RF phases, or the induction linac voltage, to match the next train.

For the 5 MHz case [31], Fig. 26a shows the phase space of muons at the end of the decay channel. The green ellipse represents the approximate bucket area of the downstream FFAG acceleration system. If, in addition, a low-frequency linear phase rotation were introduced between the decay channel and the FFAG, then muons within the magenta distorted phase space could be captured and matched into the same final bucket, yielding a better efficiency. This possible improvement has not yet been studied.

For the 44 MHz case [32, 33], Fig. 26b shows the phase space of muons before (blue) and after (green) their phase rotation. The band that is initially at a sloping angle has been rotated to a horizontal band that better fits the phase space acceptance of the following 44 MHz RF buckets. Note, however, that the small red ellipse within the phase-rotated band, which represents the approximate area of the bucket in the 88 MHz RF systems used downstream, indicates that there will be a serious inefficiency arising from this mismatch of areas.

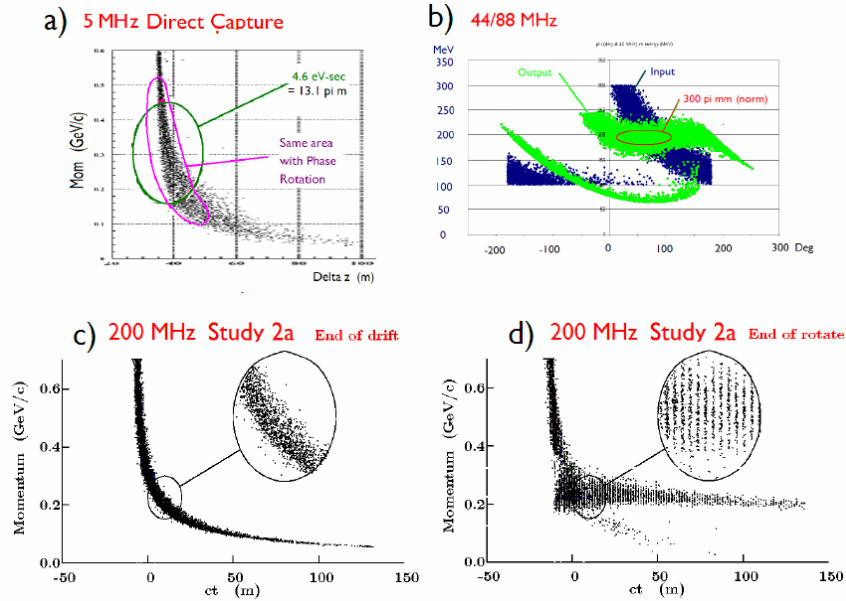

Fig. 26. Phase-space plots of phase rotation cases: a) 5 MHz before rotation with captured ellipses without rotation (green), and with rotation (magenta); b) CERN 44 MHz, before (blue) and after (green) rotation with eventual 88 MHz acceptance in red; c) and d) Study 2a before and after rotation.

For the 201 MHz case [21, 41], Figs. 26c and 26d show phase space distributions before and after the bunched beam phase rotation. Figure 26c shows the distribution prior to any RF. It is then bunched using RF with frequencies that vary along the length of the channel (see Fig. 18c), and subsequently rotated to yield the now more monoenergetic multiple bunch phase plot of Fig. 26d.

### 4.1.6  Longitudinal Capture Efficiency

Figure 27 shows the momentum ranges captured (magenta) by different phase rotation systems: a) 5 MHz without phase rotation; b) Study 2 induction linac system; c) 44 MHz rotation; d) 201 MHz bunched-beam phase rotation.

Table 12 summarizes the estimated longitudinal capture efficiencies $\eta_\parallel$, and number of muon signs captured, for different cases. The first line gives the case as presented in [31] in which the decay channel is fed directly into the first accelerating FFAG, with no matching phase rotation. The value of 39% is derived from an integration of the published momentum distribution (Fig. 27a) of initial muons, with cuts at the ±50% momentum acceptance of the first FFAG. Despite the enormous bucket area in the ring, the efficiency is not that large. The second line shows an estimated improvement that could be made with a simple linear phase rotation channel to match between initial production and the RF bucket. The third line gives the predicted performance if, in addition, a scheme is used to separate the two muon signs in the decay channel (in a bent solenoid for instance), and then inject them in opposite directions in the FFAGs. The fourth line gives the estimated efficiency for the 44 MHz case. An integration of the simulated muon momentum distribution (Fig. 27c) accepted by the 44 MHz RF bucket gives a 50% efficiency, but, as discussed above, the fraction of this bucket that is accepted in the downstream 88 MHz buckets will reduce the efficiency to only 15%. Both the fifth and eighth lines represent results using phase rotation into multiple bunches using the Neuffer scheme, which automatically captures both muon signs. The efficiency for 88 MHz is taken to be the same as that for 201 MHz. For completeness, line 6 gives the efficiency for the multi-bunch phase rotation scheme using induction linacs [24, 40]. This earlier multi-bunch scheme had a somewhat higher efficiency than the Neuffer scheme, but it does not capture both signs. Finally, line 7 gives the efficiency for the induction linac system but with (hypothetical) added charge separation and separate rotation.

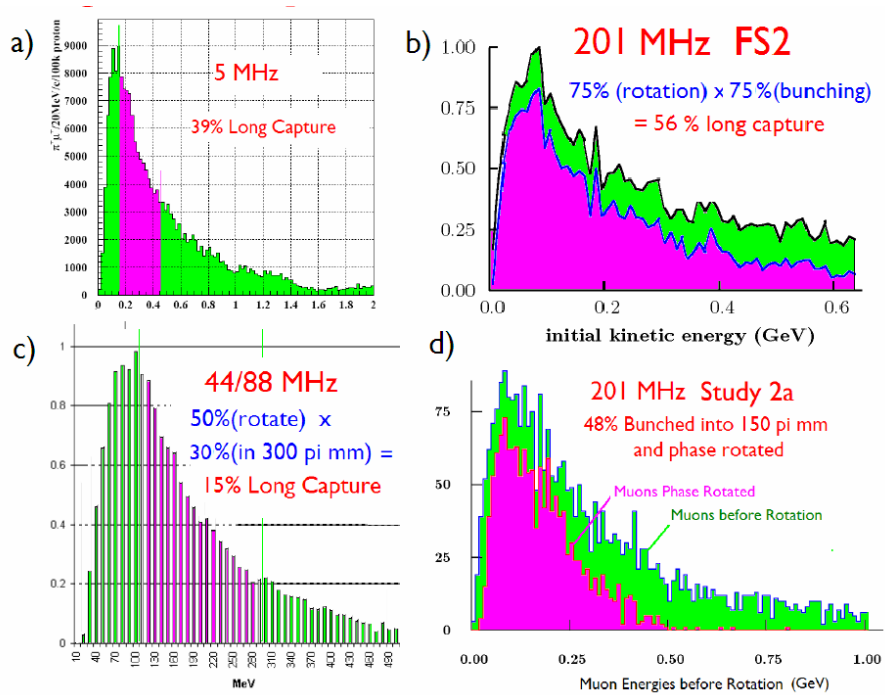

Fig. 27.  Momentum ranges captured by various phase rotation systems: a) 5 MHz without phase rotation; b) induction linac system; c) 44 MHz rotation; d) 201 MHz bunched beam phase rotation.

Table 12. Longitudinal Capture efficiencies. Parenthetical values are unpublished estimates or calculations made for this comparison study.

| $f_{RF}$ (MHz) | Case | Efficiency, $\eta_\parallel$ (%) | Signs | $\eta_\parallel \times$ signs (%) |
|---|---|---|---|---|
| 5 | No rotation | 39 | 1 | 39 |
| 5 | With rotation | (60) | 1 | (60) |
| 5 | With rotation and both signs | (60) | (2) | (120) |
| 88 | Rotation to single bunch | (15) | 1 | (15) |
| 88 | Bunched-beam rotation | (48) | 2 | (96) |
| 201 | Multibunch rotation | 56 | 1 | 56 |
| 201 | Multibunch rotation and both signs | 56 | (2) | (112) |
| 201 | Bunched-beam rotation | 48 | 2 | 96 |

It can be seen from Table 12 that the most efficient longitudinal capture uses a 5 MHz frequency, with its huge bucket area, but needs both some phase rotation and a scheme to capture both signs to compete with the higher frequency multi-bunch systems. An 88 MHz system that does not match into multiple bunches is the least efficient option. An induction linac phase rotation is marginally more efficient than the Neuffer scheme, for a single sign, but it is very expensive. Moreover, the induction linac approach needs both a scheme to separate the two muon signs and one to recombine them. The Neuffer multi-bunch method is the most attractive—it naturally captures both signs, is relatively efficient, and is cost effective.

### 4.1.7 Transverse Capture Efficiency

Figure 28 shows cooling predictions from the different reports [24, 31, 32, 40]. Figure 28a shows the emittance and transmission for cooling at 5 MHz [31] in a gas-filled FFAG. Although significant emittance reduction was obtained, this approach was not included in the baseline configuration because the transverse acceptance of the FFAG ring was sufficiently large that no more muons were captured after cooling. In such a circumstance, the only advantage of the cooling would be in cost reductions from reduced acceptance in downstream rings. We explored, by means of a simple analytic calculation of muon intensity gain versus decay loss, whether linear low frequency cooling at 5 MHz (with its larger momentum acceptance) could increase the overall efficiency. Unfortunately, because of the limit of a very low accelerating gradient (1 MV/m), we did not find any improvement.

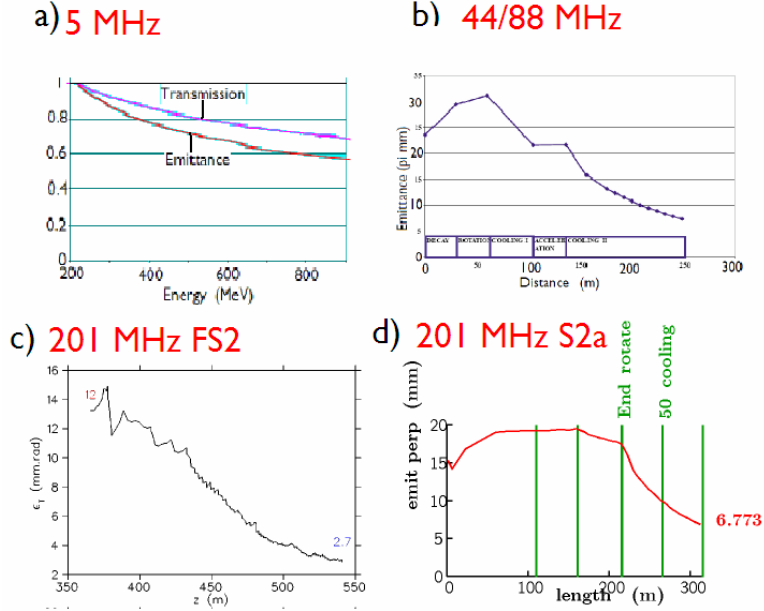

Fig. 28. Emittances *vs*. distance along cooling channel: a) 5 MHz in an FFAG; b) 44 and 88 MHz tapered system; c) Study 2 tapered system at 201 MHz; d) Study 2a non-tapered system. In a), "distance" is represented in units of the beam energy after some number of turns in the ring.

Figures 28b and 28c, respectively, show the emittances as a function of distance along the cooling channel for the CERN study at 44 and 88 MHz [32], and for Study 2 [24, 40]. In both cases, very significant cooling was required because the assumed acceptance of the downstream acceleration system was relatively low (15 $\pi$ mm-rad). Figure 28d shows the emittance *vs*. length for Study 2a [21, 41]. In this case, much less cooling was acceptable, because a larger acceptance (30 $\pi$ mm-rad, the same as in the Japanese case) was chosen.

Table 13 gives the published or estimated transverse efficiencies, $\eta_\perp$, for the different cases. The transverse efficiencies for the 5 MHz study without cooling were obtained from its overall front-end efficiency divided by the longitudinal efficiency from Table 12 above. For the other cases without cooling, we used a MARS simulation to determine the efficiency. This depends, to some extent, on the mean momentum of the muons that are selected in the longitudinal acceptance, because pions with higher overall momentum have, on average, somewhat higher transverse momenta. Efficiency values with cooling were obtained from the published values of total front-end efficiency and the longitudinal efficiencies in Table 12. The value for the CERN case was taken from [33]; Ref. [32] gives a somewhat higher number.

Table 13. Transverse capture efficiencies, with and without cooling, for the different cases. Parenthesized values are unpublished estimates or calculations made for this comparison study.

| Transv. acceptance ($\pi$ mm-rad): | | | 15 | 30 | 15 | 30 |
|---|---|---|---|---|---|---|
| Case | $p_\mu$ | $\eta_{front}$ | No cooling | | With cooling | |
| | (MeV/c) | (%) | (%) | (%) | (%) | (%) |
| 5 MHz | 300 | 15 | (7) | 18 | — | — |
| 44 + 88 MHz | 286 | 5 | (8) | (20) | 50 | — |
| FS2 | 220 | 21 | (10) | (24) | 31 | — |
| FS2a | 220 | 21 | (10) | (24) | — | 42 |

Examining Table 13, we make the following comments:

- The somewhat low acceptance in the 5 MHz case (18%) is due to the lack of cooling, despite the large acceptance of the downstream acceleration.
- The relatively higher efficiency for the tapered 44 + 88 MHz scheme (50%) compared with the similar Study 2 design (31%) is likely a result of less realistic simulations (e.g., using uniform solenoidal fields and ideal field reversals). Both are higher than the Japanese example because of efficient cooling, despite the lower acceptance (15 $\pi$ mm-rad) of the downstream acceleration system.
- The efficiency in the Study 2a case (42%) is comparable with that in FS2, reflecting the trade-off between less cooling and larger downstream acceptance (30 $\pi$ mm-rad).

### 4.1.8 Acceleration

Ignoring other losses, the acceleration efficiency, $\eta_{accel}$, is determined by decay losses. For a constant accelerating gradient, this fractional loss is given by:

$$\eta_{accel} = \frac{n_2}{n_1} = \left(\frac{E_2}{E_1}\right)^{\frac{m_\mu c^2}{c\tau G e}} \quad (6)$$

Where $\tau$ is the muon lifetime, $c$ is the velocity of light and $G$ is the accelerating gradient. Using estimates[7] of $G$ for three different frequencies we obtain the values in Table 14. Higher accelerating frequencies are clearly favored, because their higher gradients give more rapid acceleration and thus less decay loss.

### 4.1.9 Conclusions

Based on the above studies, we conclude that the preferred scheme should use:

- A proton bunch structure with bunches far enough apart to allow the muons from each bunch to be spread over a significant time interval (> 250 ns)
- Neuffer phase rotation to capture both signs into interleaved multiple bunches
- 201 MHz RF in acceleration and cooling
- An acceleration system with transverse acceptance of at least 30 $\pi$ mm-rad
- Moderate cooling

Table 14.  Acceleration efficiencies.

| | E1 (GeV) | E2 (GeV) | Gradient (MV/m) | $\eta_{accel}$ (%) |
|---|---|---|---|---|
| 5 MHz | 0.21 | 20 | 0.75–1.0 | 36–50 |
| 88 + 176 MHz | 0.20 | 20 | 1.8 | 65 |
| 201 MHz | 0.13 | 20 | 4.0 | 81 |

---

[7] Gradients given in the table are approximate "average" values, considering the fraction of acceleration system length occupied by cavities and the average accelerating phase. In [32], two different frequencies were used, with two different gradients: 4 and 10 MeV/m at 88 and 176 MHz, respectively. The value given is an "effective" one.

Such a scheme, feeding a racetrack storage ring, meets the requirement of $10^{21}$ useful muon decays per year with a 10 GeV proton source of approximately 4 MW.

## 4.2 Cooling vs. Accelerator Acceptance

There is a trade-off that can be made between the amount of cooling that must be done and the acceptance of the downstream accelerators [21, 41]. Clearly, if the accelerator acceptance were larger than the equivalent emittance of the collected muons after bunching and phase rotation, *no* cooling would be needed. This is an important concept that has significant cost implications for the Neutrino Factory design. An early study of this type is shown in Fig. 29. The line at 0.17 accepted muons per proton corresponds to the design goal for Study 2a.

The curves show the number of muons that are contained in various transverse phase space acceptances as a function of the length of the cooling channel. With the 30 $\pi$ mm-rad acceptance used in Study 2a, the length of the cooling channel must be 80 m. At the time this study was done, it appeared that increasing the accelerator acceptance to 45 $\pi$ mm-rad was possible, and would completely eliminate the need for cooling. Subsequently, it was discovered that it is very difficult to obtain transverse acceptances much larger than 30 $\pi$ mm-rad in the non-scaling FFAG accelerators, due to longitudinal phase-space distortions caused by the dependence of the time-of-flight on transverse amplitude. After these issues with the accelerator acceptance—and the ability of proposed solutions to mitigate the problem—are better understood, this question will be revisited.

## 4.3 Baseline Front-end Description

We describe here the baseline front-end design adopted for the ISS. One new feature of the adopted design is its ability to simultaneously accommodate muons of both signs. Provided the detector can handle both signs, this effectively doubles the number of useful muons per year. The schematic layout of the baseline front end is shown in Fig. 30. A summary of its main properties is given in Table 15.

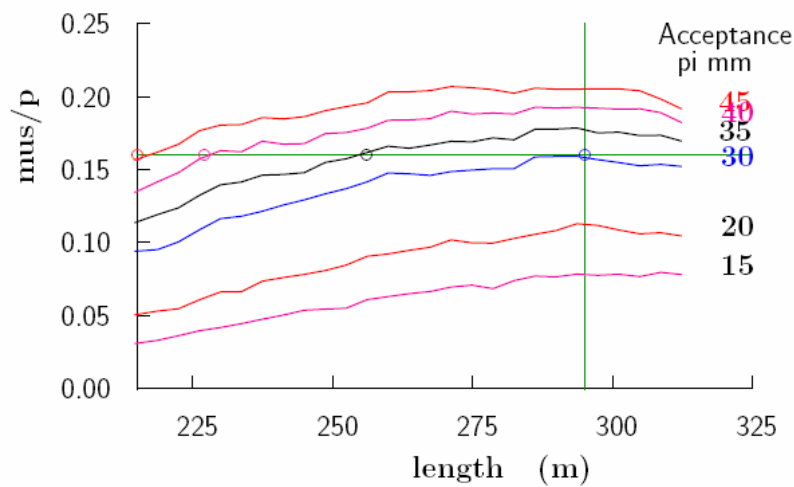

Fig. 29. Accepted number of muons per proton as a function of cooling channel length for various assumed downstream accelerator transverse acceptances.

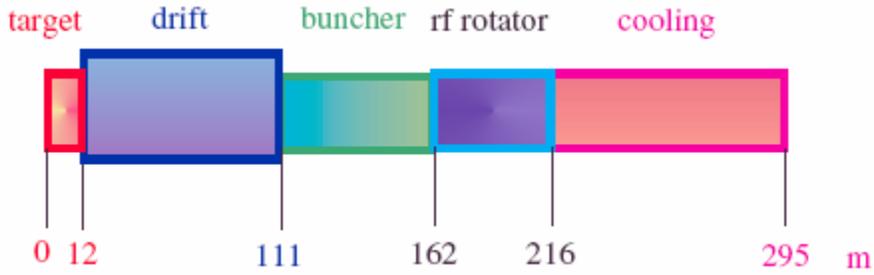

Fig. 30. Schematic layout of ISS baseline front end.

Table 15. Properties of ISS baseline front end.

| Parameter | Value |
|---|---|
| $L$ (m) | 295 |
| No. of solenoids | 460 |
| No. of RF cavities | 210 |
| Accepted muons per proton-GeV | 0.0083 |
| Transverse normalized acceptance, $\varepsilon_{TN}$ ($\pi$ mm-rad) | 7.4 |
| Helicity | 0.08 |
| $\sigma_p/p$ | 0.105 |

The baseline proton driver beam has an energy of 10 GeV. The pion collection system begins with a 20 T solenoid with 7.5 cm beam radius surrounding the target. This is followed by a 12 m long channel where the solenoid strength falls adiabatically to 1.75 T and the channel radius increases to 25 cm. There then follows a 100 m long channel where the pions decay to muons and a correlation is built up between the time and energy of the muons. This correlation in longitudinal phase space is used by the 50 m long adiabatic buncher. Bunching is accomplished with RF cavities of modest gradient, whose frequencies change as we proceed down the beam line. After bunching the beam, another set of RF cavities in the 50 m long rotator section, with higher gradients and decreasing frequencies as we proceed down the beam line, is used to rotate the beam in longitudinal phase space in order to reduce its energy spread. The final rms energy spread in this design is 10.5%. An 80 m long solenoidal focusing channel, with high-gradient 201.25 MHz RF cavities and LiH absorbers, cools the transverse normalized rms emittance from 17 $\pi$ mm-rad to about 7 $\pi$ mm-rad. This takes place at a central muon momentum of 220 MeV/c.

The cooling channel was designed to have a transverse beta function that is relatively constant with position and has a magnitude of about 80 cm. Most of the 150 cm magnetic cell length is taken up by two 50 cm long RF cavities. The cavities have a frequency of 201.25 MHz and a gradient of 15.25 MV/m. A novel aspect of this design comes from using the windows on the RF cavity as the cooling absorbers. This is possible because the near-constant beta function eliminates the need to place the absorbers at the low-beta point to prevent emittance heating. The window consists of a 1 cm thickness of LiH with a 300 μm thick layer of Be on the side facing the RF cavity field and a 25 μm thick layer of Be on the opposite side. The beryllium will, in turn, have a thin coating of TiN to prevent multipactoring. The alternating 2.8 T solenoidal field is produced with one solenoid per half cell, located between RF cavities. The channel produces a final value of $\varepsilon_{TN}$ = 7.4 $\pi$ mm-rad, more than a factor of two reduction from the initial value. The equilibrium value for a LiH absorber with an 80 cm beta function is about 5.5 $\pi$ mm-rad.

As shown in Fig. 31, the cooling channel increases the number of accepted muons by about a factor of 1.6. Normalizing to the incident 10 GeV proton beam energy on the mercury target, the figure of merit for the ISS front end is 0.0077 ± 0.0009 for the positive muons and 0.0089 ± 0.0010 for the negative muons. This efficiency is similar to the result from Study 2a [21, 41] for 24 GeV proton interactions. In addition, this channel transmits both signs of muons produced at the target. With appropriate modifications to the transport line going into the storage ring and the storage ring itself, this design would deliver both (time tagged) neutrinos and antineutrinos to the detector. The beam at the end of the cooling section consists of a train of about 80 bunches with a varying population of muons in each one.

### *4.4 Baseline Optimization Studies*

A number of front end design studies were carried out as part of the ISS program in order to do an initial optimization of the system and to identify configurations worthy of further study.

#### 4.4.1 CERN Cooling Channel

As discussed in Section 4.1, before selecting Study 2a as the baseline front end configuration, detailed comparisons were made with the front end used in the CERN Neutrino Factory studies. To permit accurate comparisons, simulations of the CERN front end [42] were made using the same initial beam distributions, simulation codes and level of detail that were used with Study 2a [43]. The original CERN front end design [30] had a 30 m decay region, a 30 m phase rotation section, 46 m of initial cooling, 32 m of acceleration, and 112 m of final cooling. The solenoidal focusing field increased from 1.8 to 5 T along the channel.

For the horn capture, we took the original CERN design, which was optimized for use with a 2.2 GeV beam. The horn design is very compact, only 1 m long with two radially nested horns extending out to a radius of 1 m. The design included a 0.5 m long drift space at the end of the horn, so the field of the first solenoid in the decay channel does not overlap the horn field. This was followed by the 44 + 88 MHz front end, except for the omission of the tapered capture solenoid. For the solenoid collection, the number of accepted positive muons ($\mu_A/\pi$) is ~0.01 for

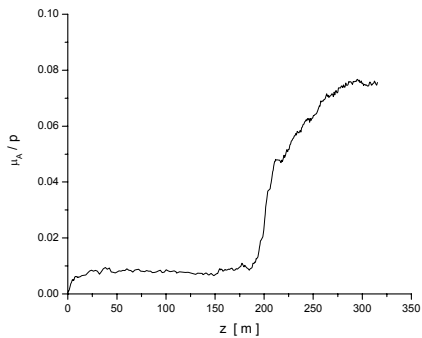

Fig. 31. Number of muons per incident proton accepted by the downstream acceleration system *vs.* longitudinal position along the ISS front end.

10 GeV protons on tantalum. Unfortunately, we found that the horn system modeled here does not work very well for the 10 GeV proton energy adopted as the ISS baseline. The compact horn design does not give sufficient focusing for a 10 GeV beam. In addition, more design work is clearly needed to transfer the beam from the 1 m radius horn to the 30 cm radius aperture of the solenoid channel. Most of the losses occur at this location.

A revised CERN design used only a single RF frequency, 88 MHz. This design had a 15 m decay region, 8 m of phase rotation, and 90 m of cooling. The solenoidal focusing field was 4 T along the whole channel. The reference particle kinetic energy was 200 MeV. No design was specified for their tapered solenoid collection system around the target, so we used the Study 2 design out to the point where the field fell to 4 T. The average RF gradient used was 4 MV/m, including transit-time factors. Simulations at CERN had shown that this revised design performed as well at 2.2 GeV as did the original 44 + 88 MHz design. Each cooling cell had a 0.5 m matching solenoid. The strength of this solenoid could be varied in order to keep the beta function in the cooling cell approximately constant.

Our initial studies used a continuous solenoid field in the channel (after the initial tapered field in the collection region). However, there is a preliminary engineering design [36] for the RF cavities that incorporates the solenoids around a neck in each cavity. The proposed solenoids are 45 cm long with an inner radius of ~15 cm. The overall length of the cavity is 90 cm. We found that this solenoid configuration produced large (~68%) modulation of the solenoid field on-axis. Large field modulation typically leads to stop-bands, so we prepared a third, low-modulation design to avoid this. In this low-modulation model, the solenoids were 60 cm long on a 90 cm period with an inner radius of 50 cm. This solenoid configuration produced only a 7% modulation on-axis, but it would be incompatible with the proposed CERN RF cavity design.

With the above modifications, the cooling channel works fairly well, but at a slower rate than in the original design simulations. The growth in the number of muons in the accelerator acceptance is shown in Fig. 32. We see continued growth in the number of accepted muons up to the end of the cooling channel, which suggests that the performance of this channel could be improved by making it longer.

We found that the all-88-MHz front end channel had significantly better performance than the 44 + 88 MHz channel. Moreover, since its total length is 113 m and its peak solenoid field is 4 T (compared with 259 m and 5 T for the 44 + 88 MHz channel), the single-frequency channel is probably less expensive to build as well. However, the number of accepted $\mu_A/\pi$ for 10 GeV protons on tantalum (0.016) is still much smaller than the corresponding number (0.097) found for the Study 2a front end. The overall figure of merit here is 0.0014 $\mu_A$/p-GeV, compared with 0.0087 $\mu_A$/p-GeV for Study 2a. This factor of 6 loss in efficiency results primarily from a problem with the mismatch between the initial longitudinal emittance of the beam at the target and the acceptance of the 88 MHz RF buckets [44]. We speculate that some of the loss could be recovered with additional optimization of the channel parameters for the higher incident beam energy, but no further work on this design has been done by its proponents.

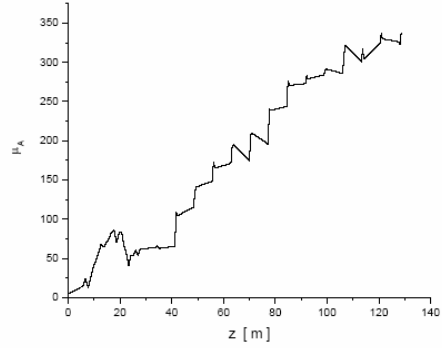

Fig. 32. Number of muons in the accelerator acceptance versus distance along the CERN channel.

### 4.4.2 Phase Rotation Optimization

Several attempts were made to understand the process of adiabatic bunching and phase rotation in more detail. The first attempt examined the behavior of the bunch centers in a simplified 1D model of phase rotation [45]. Optimization made only a small improvement in reducing the energy spread. We found that raising the RF gradient did not improve performance of this portion of the front end, and that performance degraded when the length of the phase rotation section was reduced. Changing the two reference momenta adiabatically did have a beneficial effect on performance.

In the second stage of these studies [46], the optimization program MINUIT was wrapped around the ICOOL tracking code. Five parameters were varied, with the merit function chosen as the energy spread after phase rotation. Figure 33 shows the reduction in energy spread for a set of initial particles with different energies. Once again, parameter optimization led to only small improvements in performance. We conclude that the design of the adiabatic bunching and phase rotation in the ISS front end is fairly robust and well optimized.

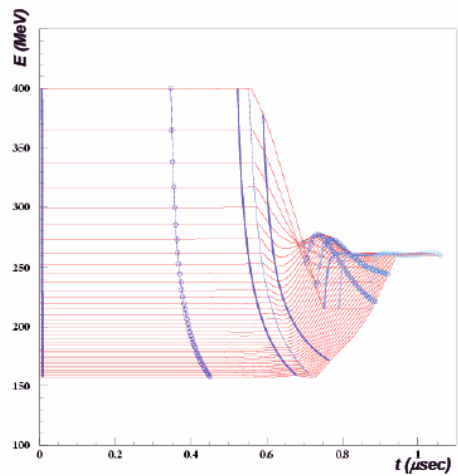

Fig. 33. Energy of a set of test particles as a function of time in the phase rotation channel.

### 4.4.3 Reduced RF Gradient

Based on studies with 805 MHz cavities, we know that the magnitude and direction of a magnetic field can have a substantial effect on the maximum achievable RF cavity gradient. Similar experimental studies with 201 MHz cavities should take place within the next year. With this in mind, we decided to investigate the performance sensitivity to operating with reduced RF gradient. The baseline cooling channel assumes 201 MHz cavities with a gradient of 15 MV/m in a 2.8 T solenoid field. Here, we examined [47] the effect of assuming that the maximum achievable gradient was only 10 MV/m. There was a net loss of 20% in the number of accepted muons after the RF phases and absorber thicknesses were readjusted. We also looked into the case where cavities are produced with a distribution of gradients. It was found that the best performance resulted from putting the highest gradient cavities at the beginning of the channel. If as few as 12 full-gradient cavities are available, the baseline front end performance can be maintained.

### 4.4.4 RF Cavity Failures

As the RF cavities in the ISS baseline cooling channel must operate at a high gradient, one concern is the effect on the front end performance if one of the cavities becomes inoperable. Studies showed that the random loss of one cavity caused only a 3% drop in the number of accepted muons [48].

### 4.4.5 Curved RF Windows

The baseline RF cavity design for the ISS cooling channel uses curved end windows. The curvature causes a distortion of the cylindrical pillbox RF fields normally assumed in the simulations. This was studied [49] by importing SUPERFISH models of the cavity with curved windows into ICOOL. Because it is the limiting aperture in the cooling channel, reducing the window radius from 25 cm, as assumed in Study 2a, to 21 cm as designed for MICE, decreases the μ/p into the accelerator acceptance by ~6%. Changing the flat RF windows assumed in Study 2a to the stronger curved windows that will be used in MICE had no statistically significant effect on the performance.

### 4.4.6 Gradient and Field Errors

Studies were made of the effects on performance of errors in the magnitude of the RF gradient and the solenoid field strength [47]. Preliminary results show that the fractional change in performance is approximately equal to the fractional errors in the gradient or field strength.

### 4.4.7 Alternative Designs

Some alternative front end designs were considered that differ in significant ways from the ISS baseline. Some of these ideas could potentially be incorporated into backup designs if the ISS baseline system encounters any unexpected difficulties.

#### 4.4.7.1 Phase Rotation with a Scaling FFAG

The PRISM project uses a scaling FFAG to produce a small energy spread by longitudinal phase space rotation. An OPERA field map of the PRISM magnetic field was used with ICOOL to model this ring [50]. The RF cavities were modeled with a simple sawtooth waveform. Closed orbits were found and sample particles were used to estimate the dynamic aperture. Tracking was also done with Gaussian beams. Figure 34 shows the decrease of the energy spread in the beam with time. Clear phase rotation is seen over seven turns in the ring. To make it suitable to serve as part of the ISS front end, the field in the FFAG was scaled from the 68 MeV/c used in PRISM to the 200 MeV/c required for a Neutrino Factory. Unfortunately, simulations to date with the scaled ring have not yet demonstrated good phase rotation.

### 4.4.7.2 Low-Frequency Phase Rotation

Studies [51] were made of an alternative 31 MHz phase rotation system that could simultaneously capture both signs of muon charge. This channel first separates the signs with on-peak RF cavities. Then a drift is used to get a separation in time. Next, the bunches are placed on opposite sides of two adjacent wave troughs. At this point, further phase rotation operates on both signs simultaneously. The system works fairly well, although at present the overall efficiency is smaller than Study 2a and the channel is longer. It is expected that improvements could be made using higher harmonics in the RF system.

### 4.4.7.3 Gas-Filled Cooling Channel

An alternative front end configuration was investigated that did the ionization cooling in the phase rotation channel [52]. High pressure hydrogen gas was used as the absorber. This approach might be less expensive, due either to the reduced channel length or to the presumed higher gradients available in gas-filled RF cavities. Simulations with 150 atm of hydrogen and 24 MV/m gradients in the cavities achieved similar performance to a simplified model of Study 2a. This can be seen from the light blue curve in Fig. 35. The effect of the required high-pressure beam windows was shown to be acceptable.

Other simulations showed that the length of the baseline phase rotator could be reduced from 54 m to 27 m with a reduction of only 10% in the number of accepted muons. However, replacing the distributed gas absorber with thicker LiH or beryllium windows in the combined phase-rotator–cooler gave significantly worse performance.

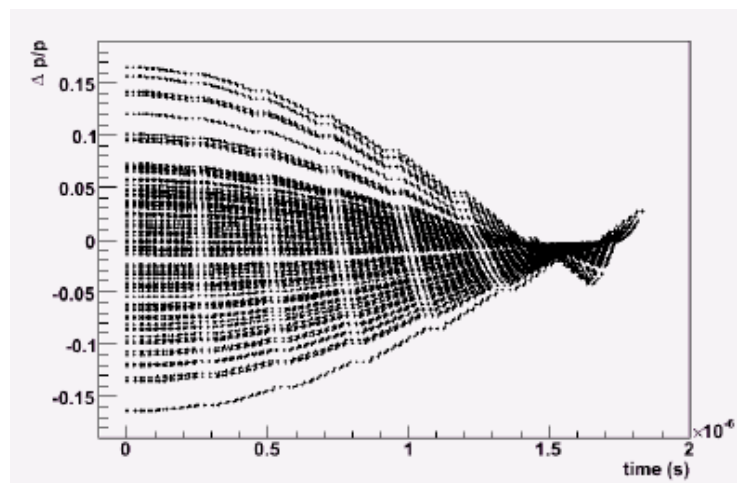

Fig. 34. Energy spread of Gaussian beam versus time in the simulation model of PRISM.

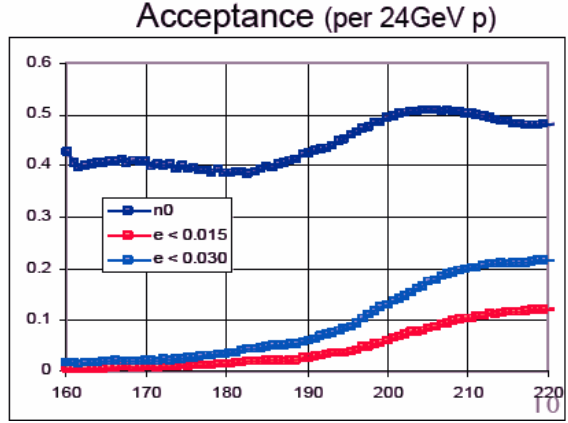

Fig. 35. Accepted number of muons per incident proton as a function of distance in the gas-filled cooling channel. The upper (dark blue) curve is the total number of μ/p, the middle (light blue) curve is μ/p into the ISS acceptance, and the bottom (red) curve is μ/p into the Study 2 acceptance.

### 4.4.7.4 Guggenheim Cooling Channel

Much of the front end muon loss occurs because of particles falling out of the 201 MHz RF buckets. If longitudinal cooling were available in the front end, these losses could be reduced and more neutrinos would be available from the decay ring (or, alternatively, the required proton driver power could be reduced). The Guggenheim cooling channel transforms the well-studied RFOFO cooling ring into a helix [53]. This avoids the challenging injection problem of getting a very large emittance beam into a small ring, and allows tapering of the channel parameters for optimal performance, but at the cost of no longer accommodating simultaneous transport of both muon signs. In initial studies, the field map for the RFOFO ring was transformed into a helical geometry with a 3 m pitch. Simulations without taking into account any iron shielding showed similar cooling performance to the RFOFO ring. Work is proceeding toward making a more realistic model of the field and matching to the front end channel.

## 5. Acceleration System

The goal of the acceleration system is to increase the beam kinetic energy from 138 MeV (the average kinetic energy in the cooling section) to a final energy in the range of 20–50 GeV. The layout described here will accelerate to 25 GeV, with an option of doubling that final energy to 50 GeV.

The design of the acceleration system is based on considerations of both cost and performance, taking into account the penalty associated with decays and other losses. The chosen approach also aims to minimize transverse and longitudinal emittance growth during acceleration. To minimize the effects of muon decay, particles must be accelerated as rapidly as possible. This is made more difficult by the fact that the beam sizes, both transverse and longitudinal, are very large. For these studies, the transverse normalized acceptance is chosen to be 30 $\pi$ mm-rad, and the longitudinal normalized acceptance is 150 mm. The transverse normalized acceptance is defined to be $a^2 p/\beta mc$, where $a$ is the maximum half-aperture at a given location, $\beta$ is the Courant-Snyder beta function at that point, $p$ is the total momentum, $m$ is the muon mass, and $c$ is the speed of light. The longitudinal normalized acceptance is defined for an upright ellipse to be $\Delta t \Delta E/mc$, where $\Delta t$ is the maximum half-width in time of the beam, and $\Delta E$ is the maximum half-width in energy of the beam.

The types of subsystems and their sequence are similar to what was proposed in [21]. However, significant changes in the details have occurred, in particular the energies for the transitions between the subsystems.

## 5.1 Overall Scenario

The acceleration system consists of several different types of subsystems. The choice of where to use which type of subsystem is governed by beam dynamics and cost considerations. At this point, a detailed cost optimization has not been performed and many of the beam dynamics issues are still being studied. Thus, the scenario chosen here is based on initial estimates of machine performance and on past experience with the cost behavior of these systems.

High average gradients are necessary to minimize the amount of muon decay. Superconducting cavities are used to keep the RF power required to achieve these high gradients modest. Since the RF systems (cavities, cryostats, RF power systems, and associated cryogenic systems) tend to be the most expensive component of the acceleration systems, and since RF cavities generally operate most economically at or near their highest achievable gradient, minimizing cost implies minimizing the number of RF cavities used. To reduce the required number of RF cavities, we chose designs where the beam makes multiple passes through the RF cavities. The choice of which subsystem to use is based primarily on the number of passes through the cavities it can accommodate.

Figure 36 shows a diagram of the entire acceleration system. The following subsections will explain the different types of subsystems, why they were chosen, and the reasons behind the particular choices of energy transition points.

## 5.2 Pre-Acceleration Linac

We begin acceleration with a linac. This avoids problems found in a recirculating accelerator with large beam sizes and large relative energy spreads (in particular, the variation of the velocity with energy, which means that if the phase relationship between cavities in the RLA linac is correct for the final linac pass, it is incorrect for the initial pass). The linac is used to accelerate to a point where such effects can be handled in the RLA. It comprises three different styles of cryo-modules, having increased cell length as the beam energy increases. Transverse beam envelopes and longitudinal profiles along the linac are shown in Fig. 37.

The initial longitudinal acceptance of the linear accelerator is chosen to be $2.5\sigma$, i.e., $\Delta p/p = \pm 0.17$ and RF pulse length $\Delta \phi = \pm 92°$. To perform adiabatic bunching, the RF phase of the cavities is shifted by 72° at the beginning of the pre-accelerator and gradually changed to zero by the linac end. In the first half of the linac, when the beam is still not sufficiently relativistic, the offset causes synchrotron motion, allowing bunch compression in both length and momentum spread to $\Delta p/p = \pm 0.07$ and $\Delta \phi = \pm 29°$. The synchrotron motion also suppresses the sag in acceleration for the bunch head and tail. Figure 38 shows how the initially elliptical boundary of the bunch longitudinal phase space will be transformed by the linac.

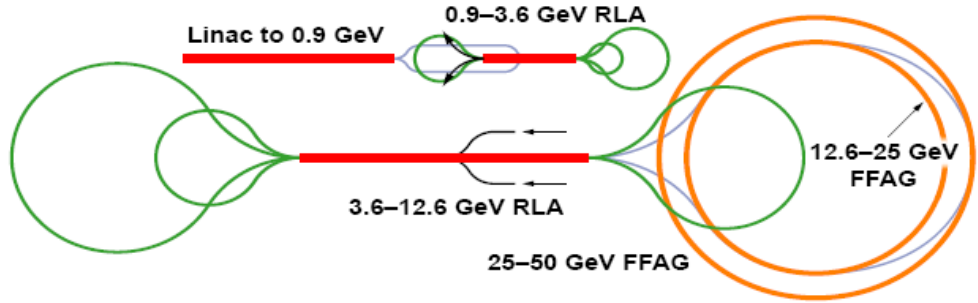

Fig. 36. Layout of the acceleration system.

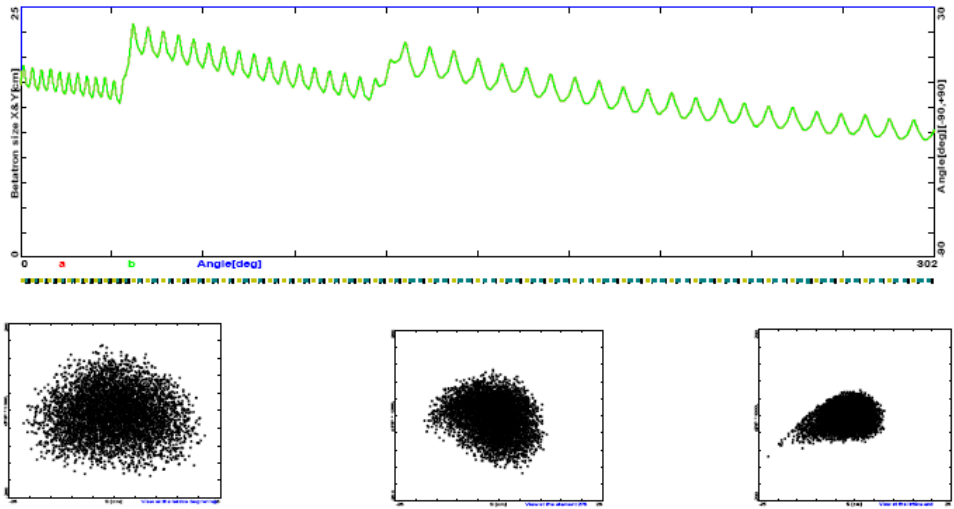

Fig. 37. Top, transverse beam envelopes in the pre-accelerator linac, which has uniform periodic focusing with three styles of cryo-modules. Below, longitudinal phase-space: before, halfway through, and at the end of acceleration, as obtained by particle tracking.

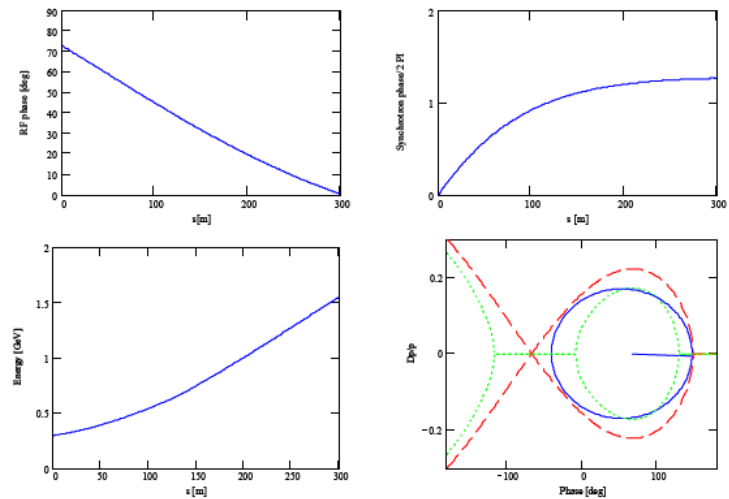

Fig. 38. Individual cavity phasing along the linac and the resulting synchrotron motion. The energy profile and the longitudinal beam boundary (solid line) inside a separatrix (dashed line) are shown at the beginning of the linac ($\Delta p/p = \pm 0.17$ or $\Delta\phi = \pm 92°$.).

We choose the final energy of the pre-accelerator linac to be 0.9 GeV. While a detailed parameter optimization has not been completed, this energy is expected to be the lowest energy that will keep the phase slip in the first RLA tolerable in terms of the velocity difference between the first and last linac passes.

## *5.3 Recirculating Linear Accelerators*

Recirculating linear accelerators (RLAs) are machines that take one or more linacs and connect them by a series of arcs. After each pass though the linac, the beam enters a different arc, which will transport it to the next linac or the next pass through the same linac. The switchyard, where the beam from the linac is transported into each individual arc, uses fixed-field magnets. Because of the nonzero energy spread in the beam, the nonzero transverse beam size, the space required for magnet coils, and other considerations, the number of separate arcs the beam can be directed into is typically limited to 4 or 5. This, in turn, limits the number of passes through the cavities that an RLA can achieve.

One way to increase the efficiency of an RLA is to change its geometry. Figure 39 shows two different layouts: a racetrack layout and a dogbone layout. The racetrack layout is, in principle, more straightforward to design and build: the arcs bend in only one direction, and there is no need to introduce vertical bending to avoid beam line crossings that occur when one tries to minimize arc length. The dogbone geometry, on the other hand, is more efficient. In particular, since the energy separation at the switchyard effectively limits the number of passes the beam can make through the linac, the dogbone layout allows twice as many passes through the linac as the racetrack layout, and is thus preferred.

The baseline design has two dogbone RLAs following the pre-acceleration linac, as indicated in Fig. 40. We chose the maximum RLA energy to be 12.6 GeV. Using two RLAs allows a lower injection energy in the first one than if only one RLA were utilized. Furthermore, it potentially increases the amount of synchrotron oscillation in the RLAs, reducing the effects of the time-of-flight dependence on transverse amplitude in the RLA linacs (see below) and differential beam loading down the bunch train. All subsystems in Fig. 40 accommodate simultaneous acceleration of muons of both signs.

Selection of a higher number of passes in each RLA (4.5 vs. 3.5 in Study 2a) for the baseline scheme is driven by recent successful studies of FODO-based lattices, which are more suitable (compared with triplet focusing) to accommodate a large number of passes in a dogbone configuration. The new focusing structure (described below) offers a well balanced multi-pass linac optics as well as uniform beta matching to the droplet arcs. Furthermore, the FODO structure can still support a compact spreader and recombiner optics and a uniform dispersion flip for the droplet arcs, as will be illustrated below.

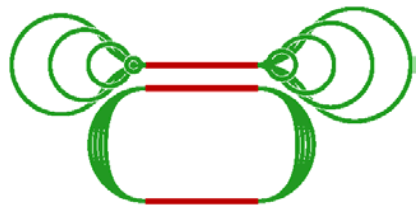

Fig. 39. RLA geometries: dogbone layout (above), racetrack layout (below).

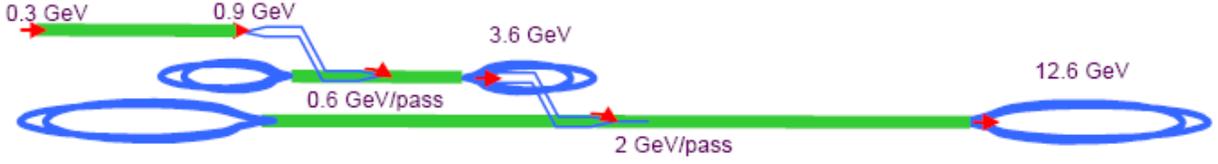

Fig. 40. Layout of two-step dogbone RLA complex. For compactness, all three subsystems (pre-accelerator, dogbone I and dogbone II) are stacked up vertically; $\mu^\pm$ beam transfer between the accelerator components is facilitated by the vertical double chicane (see text).

The energy range for each dogbone RLA was chosen to give similar ratios of top-to-injection energies: namely 3.5 for dogbone I and 4.0 for dogbone II. Furthermore, the injection energies for both RLAs were chosen so that a tolerable level of RF phase slippage along a given length of linac can be maintained. A simple calculation of the phase slippage of a muon injected with initial energy $E_0$ and accelerated by $\Delta E$ in a linac of length $L$, where uniformly spaced RF cavities are phased for a speed-of-light particle, was carried out using the following cavity-to-cavity iterative algorithm for the phase-energy vector:

$$\begin{pmatrix} \phi b_{k,i+1} \\ Eb_{k,i+1} \end{pmatrix} := \begin{bmatrix} \phi b_{k,i} + \frac{h}{\lambda} \cdot 360 \left[ \frac{1}{2} \left( \frac{m_\mu}{Eb_{k,i}} \right)^2 \right] \\ Eb_{k,i} + h \cdot \Delta E_k \end{bmatrix} \qquad (7a)$$

where

$$h := \frac{L_{linac}}{N_{cav}} \qquad \lambda := \frac{c}{f_0} \qquad k := 0..4 \qquad i := 0..1 N_{cav} - 1 \qquad (7b)$$

The resulting phase slippage profile along the multi-pass linacs is illustrated in Fig. 41 for both dogbones I and II. The injection energy to dogbone I, set at 0.9 GeV, results in a still manageable phase slip of about 40° for the initial 'half-pass' through the linac. The corresponding value of phase slip for dogbone II is much lower, about 10°.

Initial bunch length and energy spread are still too large at the RLA input and further compression is required in the course of acceleration. To accomplish this, the beam is accelerated off-crest with nonzero $M_{56}$ (momentum compaction) in the droplet arcs. This induces synchrotron motion, which suppresses the longitudinal emittance growth related to non-linearity of accelerating voltage. Without synchrotron motion, the minimum beam energy spread would be determined by nonlinearity of the RF voltage along the bunch length, and would be equal to $(1 - \cos\phi) \approx 9\%$ for a bunch length of $\phi = 30°$. The synchrotron motion within the bunch averages the total energy gain of tail particles to the energy gain of particles in the core. We designed for the same values of $M_{56}$ for all droplet arcs—the optimum value is about 5 m—whereas optimal detuning of RF phase from the on-crest position is different for different arcs (see 'gang phases' for various passes listed in Fig. 41).

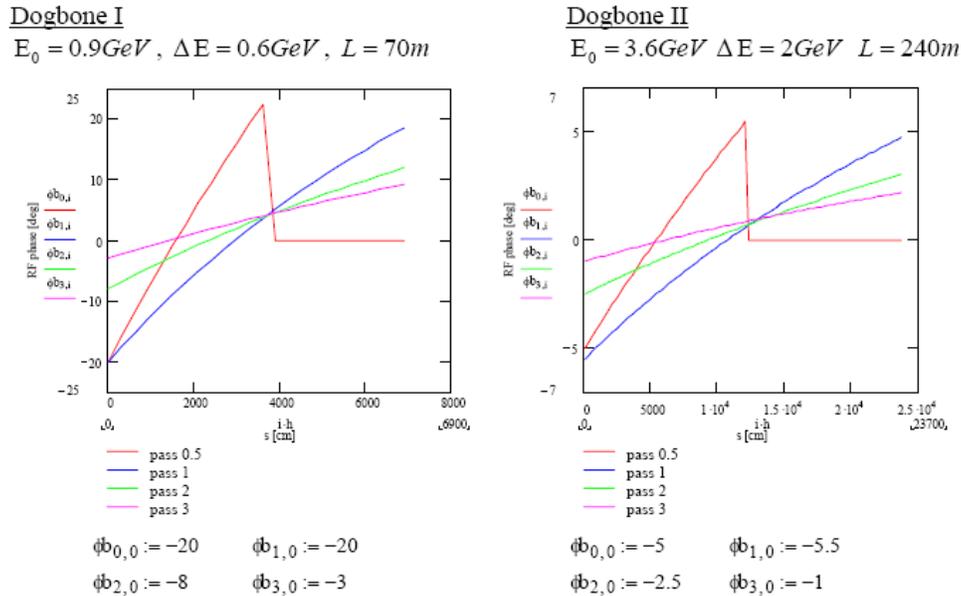

Fig. 41. RF phase slippage along the multi-pass linacs; initial 'gang phases' for each pass (listed at the bottom of the plots in RF degrees) were chosen for optimum longitudinal bunch compression in each linac-arc segment.

### 5.3.1 Multi-pass Linac Optics – FODO versus Triplet Focusing

Two styles of focusing (FODO and triplet) were considered as a base for building RLA lattices. The requirement of quasi-periodic focusing throughout the entire beam line imposes the constraint that a consistent style be used for all lattice segments (multi-pass linac and recirculation arcs). Features of both focusing styles are summarized in Fig. 42. Compared with triplet optics, the advantages of the FODO optics include:

- much weaker quadrupoles (1/3 the strength of triplet quadrupoles)
- lower integrated quadrupole length
- easier chromaticity correction

On the other hand, triplet focusing offers some advantages over the FODO design:
- longer straight sections
- lower vertical beta functions
- more uniform beta functions and dispersion

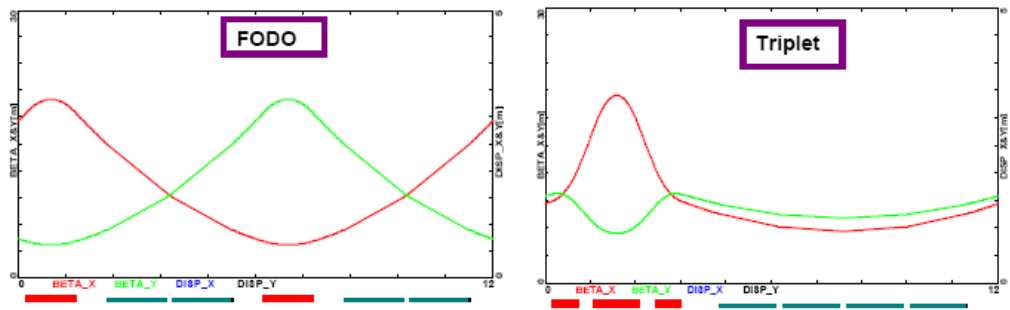

Fig. 42. Comparison of FODO and triplet focusing structures. Both designs use the same cell length (12 m) and betatron phase advance per cell (90°).

The key element of the transverse beam dynamics in a multi-pass dogbone RLA is an appropriate choice of multi-pass linac optics. The focusing profile along the linac (quadrupole gradients) need to be set (and remain constant), so that one can transport (provide adequate transverse focusing for a given aperture) multiple pass beams over a vast energy range. Obviously, one would like to optimize the focusing profile to accommodate the maximum number of passes through the RLA. In addition, the requirement of simultaneous acceleration of both $\mu^\pm$ species imposes mirror symmetry of the droplet arc optics (the two species move in the opposite directions through the arcs). This, in turn, puts a constraint on the exit and entrance Twiss functions for two consecutive linac passes, namely $\beta_{\text{out}}^n = \beta_{\text{in}}^{n+1}$ and $\alpha_{\text{out}}^n = -\alpha_{\text{in}}^{n+1}$, where $n = 0$, 1, 2... is the pass index.

We examined both styles of focusing (triplet and FODO) to design the optimum multi-pass linac optics for the dogbone RLA. The example presented below describes a dogbone based on a 2 GeV per pass linac (240 m long) with an injection energy of 2 GeV. Since the beam is traversing the linac in both directions throughout the course of acceleration, we choose a flat focusing profile for the entire linac, e.g., the quadrupoles in all cells are set to the same gradient, corresponding to 90° phase advance per cell determined for the lowest (injection) energy. There is no scaling up with energy for the quadrupole gradients along the linac. Multi-pass optics have been developed for both the triplet (Fig. 43) and FODO (Figs. 44 and 45) linacs. One can see immediately the superiority of the FODO structure, which supports twice as many passes through the dogbone RLA (6 vs. 3).

### 5.3.2 Droplet Arcs

In a dogbone RLA one needs to separate different energy beams coming out of the linac and to direct them into appropriate droplet arcs for recirculation. For many practical reasons, horizontal rather than vertical beam separation was chosen. Rather than suppressing horizontal dispersion created by the spreader, it is smoothly matched to the horizontal dispersion of the outward 60° arc. Then, by an appropriate pattern of removed dipoles in three transition cells, one 'flips' the dispersion for the inward bending 300° arc, etc. The entire droplet arc architecture is based on 90° phase advance cells with periodic beta functions. The lattice building blocks, along with the droplet arc footprint, are illustrated in Fig. 46.

The resulting droplet arc optics based on triplet focusing is illustrated in Fig. 47. One can easily estimate the momentum compaction of the arc as follows:

$$M_{56} = -\int \frac{D_x}{\rho} ds = -D_x^{dip} \int d\left(\frac{s}{\rho}\right) = -D_x^{dip} \int d\theta_{rad} = -D_x^{dip} \times \theta_{rad}^{tot}$$

$$M_{56} = -\frac{5}{3}\pi \times 1.2 \, m = -6.3 \, m$$

(8)

Similarly, one can design droplet arc optics (analogous to the one illustrated in Fig. 47) using FODO cells as the basic building blocks. The spreader optics for both styles of focusing is illustrated in Fig. 48.

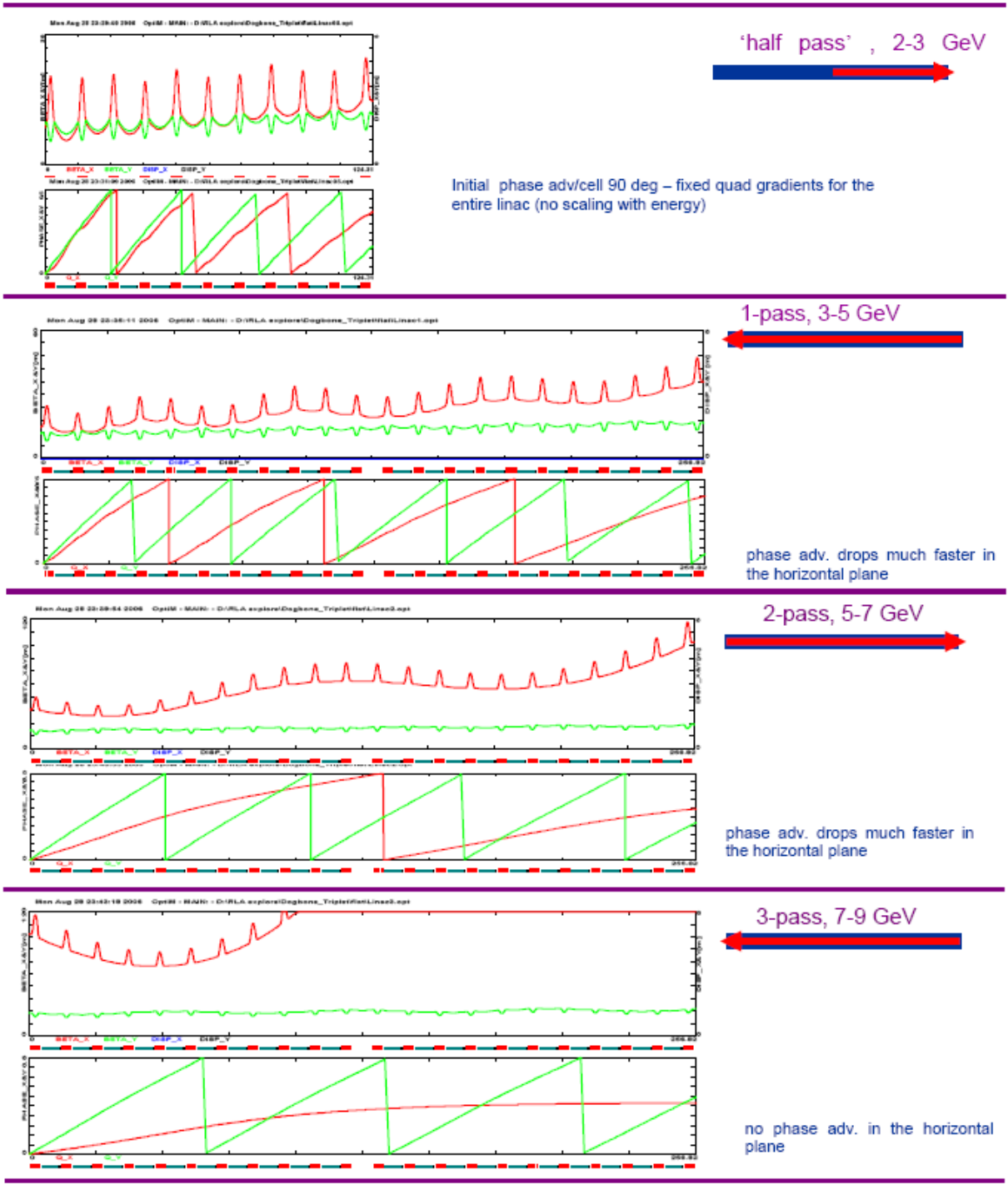

Fig. 43. Triplet based multi-pass linac optics. Quadrupoles in all cells are set to the same gradient, corresponding to 90° phase advance per cell at 2 GeV. Intrinsic to triplet focusing, the asymmetry between the horizontal and vertical planes (middle quadrupole focuses horizontally, outer quadrupoles focus vertically) manifests itself via a rapid phase advance loss in the horizontal plane for higher passes; already by the third pass horizontal focusing is almost lost, resulting in catastrophic beam blow-up.

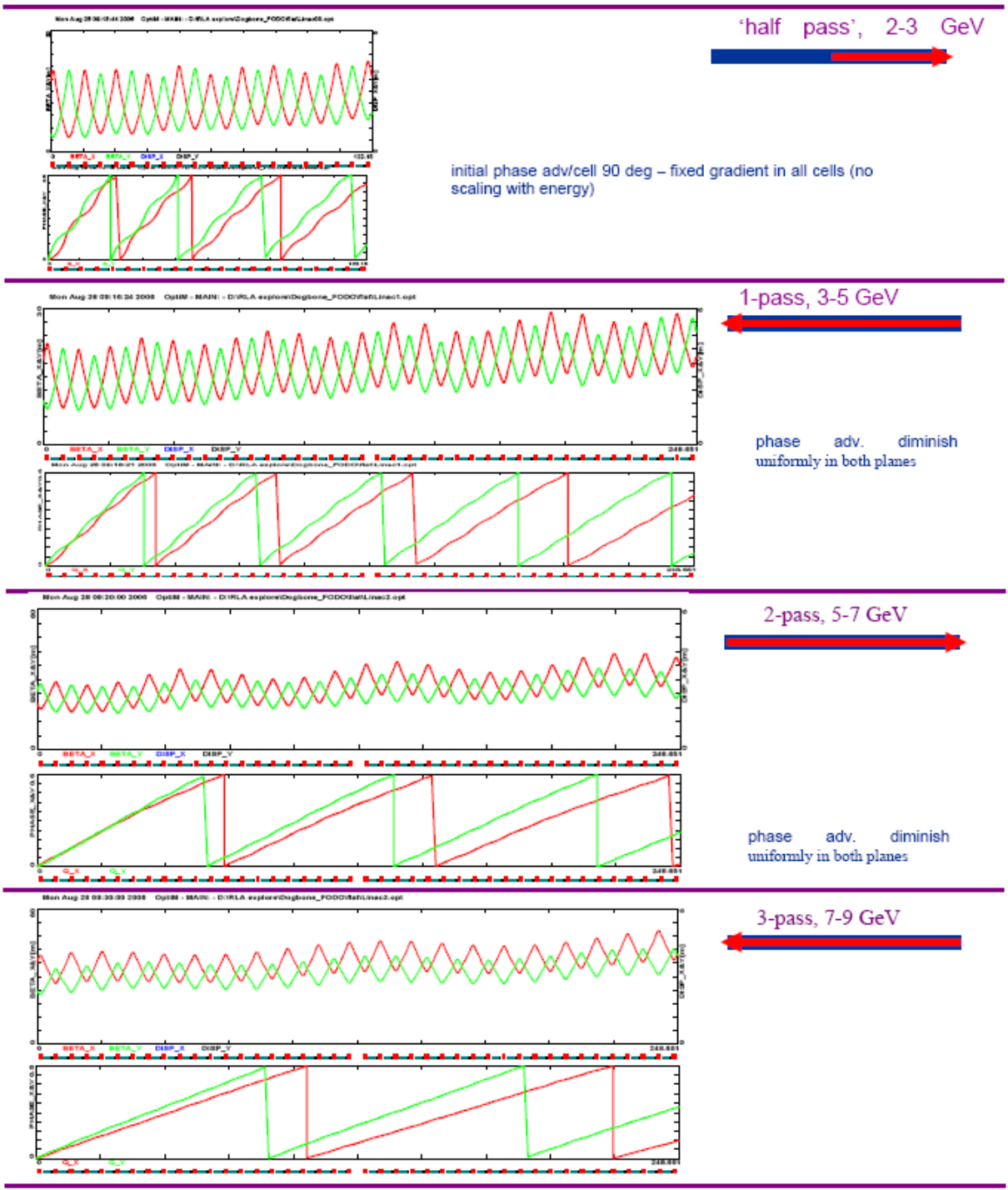

Fig. 44. FODO based multi-pass linac optics for passes 1–3. Quadrupoles in all cells are set to the same gradient, corresponding to 90° phase advance per cell at 2 GeV. Intrinsic to the FODO focusing symmetry between the horizontal and vertical planes, the betatron phase advances gradually diminish uniformly in both planes. The resulting linac optics is well balanced in terms of Twiss functions and beam envelopes, and there is sufficient phase advance for up to six passes.

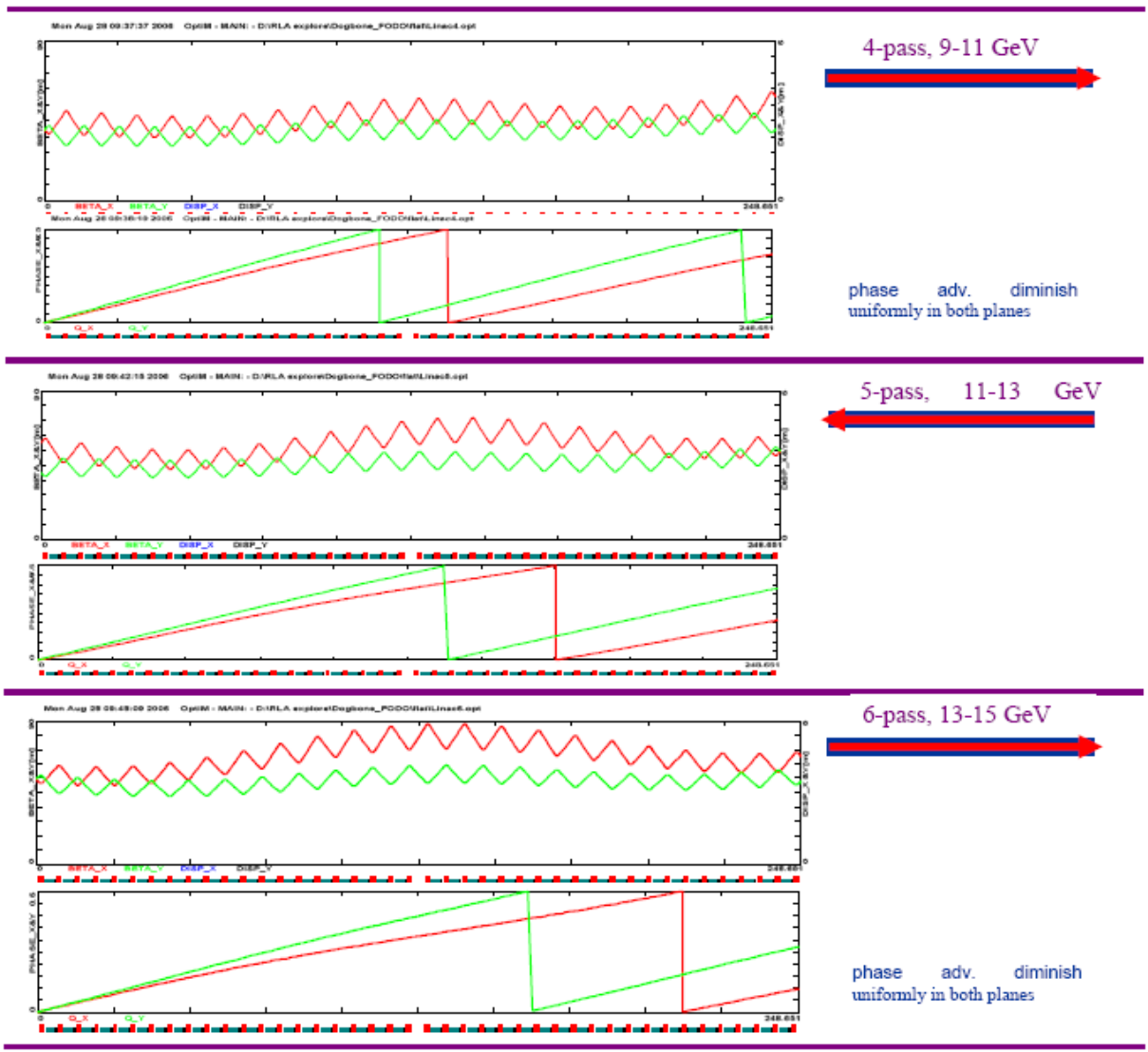

Fig. 45. FODO based multi-pass linac optics for passes 4–6. Quadrupoles in all cells are set to the same gradient, corresponding to 90° phase advance per cell at 2 GeV. Intrinsic to the FODO focusing symmetry between the horizontal and vertical planes, the betatron phase advances gradually diminish uniformly in both planes. The resulting linac optics is well balanced in terms of Twiss functions and beam envelopes, and there is sufficient phase advance for up to six passes.

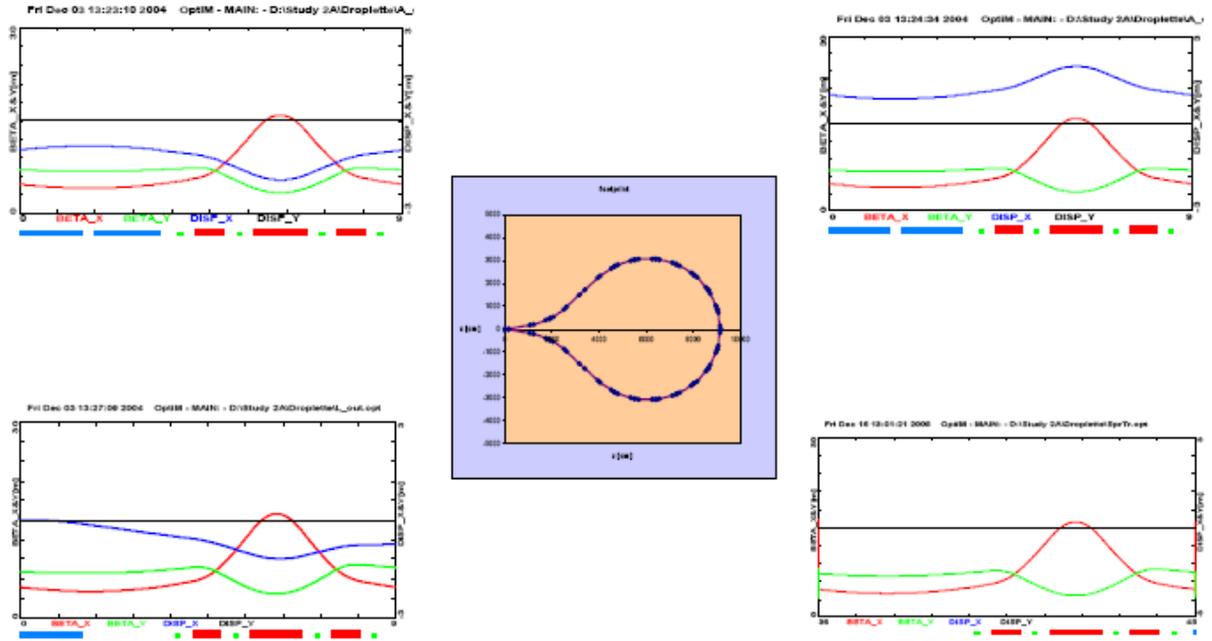

Fig. 46. The lattice building blocks: triplet 90° phase advance cells; inward and outward cells, missing dipole empty cells.

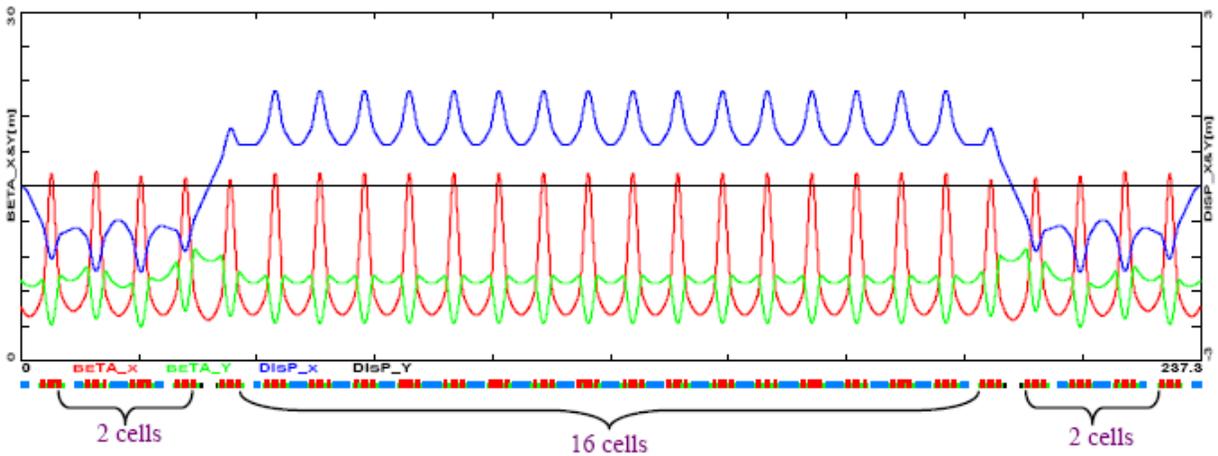

Fig. 47. Droplet arc optics, showing uniform periodicity of beta functions and dispersion.

### 5.3.3 Injection Double Chicane

To transfer both $\mu^+$ and $\mu^-$ species from one RLA to the other, which is located at a different vertical elevation, we designed a compact double chicane based on a periodic 90° phase advance cell (in either triplet or FODO style). Each leg of the chicane involves four horizontal and two vertical bending magnets, forming a double achromat in the horizontal and vertical planes, while preserving the periodicity of beta functions. The layout and Twiss functions of the double chicane are illustrated in Fig. 49.

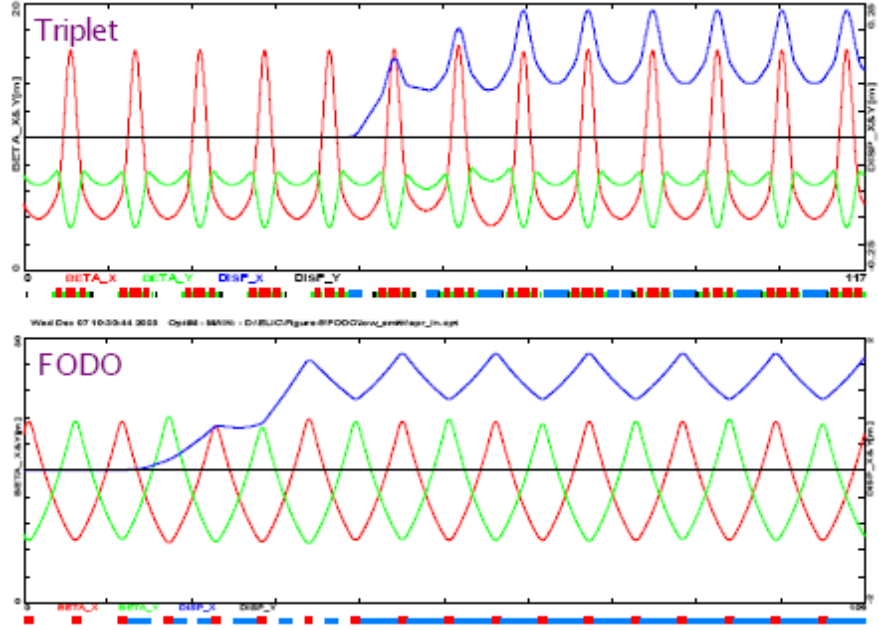

Fig. 48. Spreader optics showing uniform periodicity of beta functions and dispersion. These optics apply to either the FODO or the triplet design.

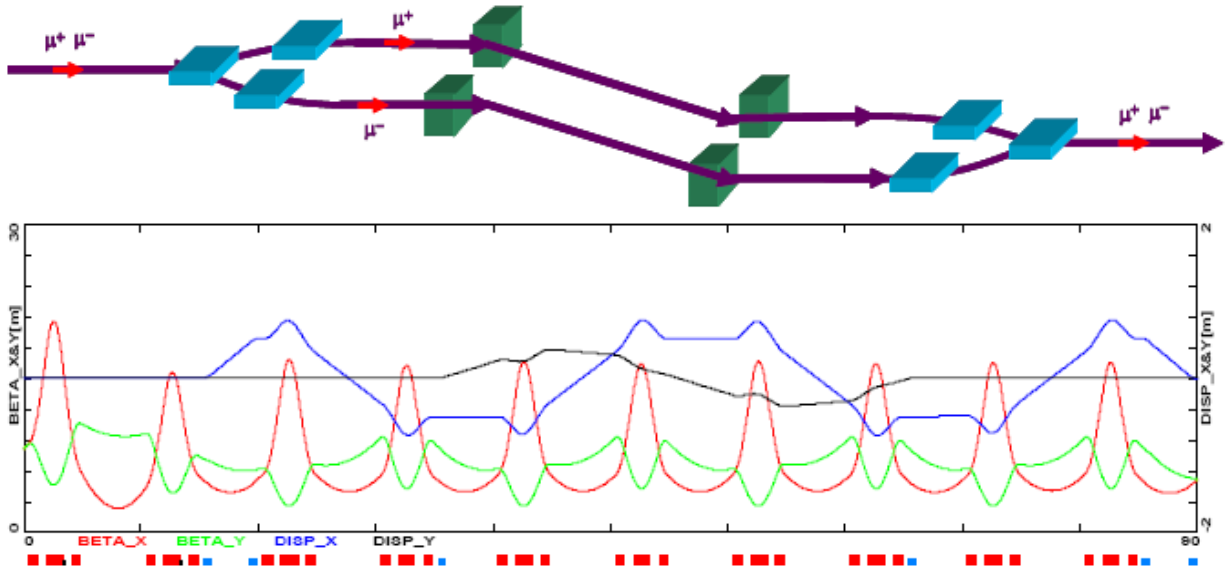

Fig. 49. Layout and optics of the injection double chicane.

### 5.3.4 Magnet Error Tolerances

For completeness, all lattices, both linacs and droplet arcs, were checked for error sensitivities. First, lattice sensitivity to random misalignment errors was studied for a droplet arc via DIMAD, assuming 1 mm (rms) quadrupole misalignments in $x$ and $y$. The resulting orbit deviation is illustrated in Fig. 50. The level of a few mm orbit drift between adjacent cells can easily be corrected by a pair of 0.002 T-m correctors located at every girder. Furthermore, the lattices were tested for magnetic field error tolerance. By design, one can tolerate some level (say, 10%) of arc-to-arc betatron mismatch due to the focusing errors, $\delta\phi_1$ (quadrupole gradient errors and dipole body gradient), which can partially be compensated by dedicated matching quadrupoles. The resulting focusing error tolerance has been evaluated for the arc 2 lattice segment from linac 2 using Eq. (9). A 10% betatron mismatch requires quadrupole errors of 0.001, which should be achievable.

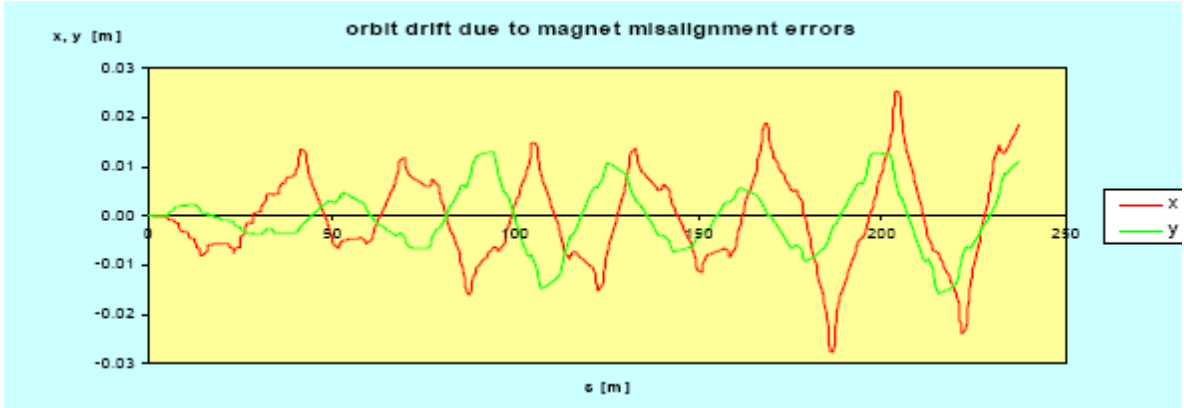

Fig. 50. Orbit deviation due to magnet misalignment errors.

$$\left(\frac{\sigma_\varepsilon}{\varepsilon}\right)_{mis} = \sqrt{\frac{1}{2}\sum_{n=1}^{N}(\beta_n \delta\phi_1)^2} = \sqrt{\frac{1}{2}\Delta\phi_1^2 \sum_{n=1}^{N}(\beta_n)_{quad}^2 + \frac{1}{2}\delta\phi_1^2 \sum_{n=1}^{N}(\beta_n)_{dipole}^2} \tag{9}$$

### *5.4 Fixed Field Alternating Gradient Accelerators (FFAGs)*

To avoid the limitations of the RLA switchyard, it is possible to utilize a single arc for all beam energies. This is what is known as a fixed-field alternating-gradient (FFAG) accelerator. All FFAGs consist of a sequence of simple, identical cells with RF cavities in most of them. The design of the cell determines the type of the FFAG and the method by which beams must be accelerated. For this study, both scaling and non-scaling FFAGs were considered.

#### 5.4.1 Scaling FFAGs

Scaling FFAGs are the original type of FFAG that was first described and built in the 1950s [54–56]. A design study for a Neutrino Factory based solely on FFAGs for muon acceleration was completed in 2001 [31]. We have chosen not to use scaling designs in the ISS baseline configuration for two reasons.

1. Scaling FFAGs generally require relatively low frequency RF cavities (of the order of 15 MHz) in order to accelerate muons, due to the relatively large time-of-flight variation with energy in these machines. This would require the earlier capture systems to use the same low frequency RF, which significantly decreases the capture efficiency of the machine.

2. Scaling FFAGs require relatively large aperture high-field superconducting magnets, which are expensive.

At lower energies, it may be possible overcome these difficulties of using scaling FFAGs.

### 5.4.2 Linear Non-Scaling FFAGs

Linear non-scaling FFAGs [57, 58] attempt to address the two main difficulties of scaling FFAGs (large aperture and large time-of-flight variation with energy) by addressing the underlying reason for these problems—in a linear non-scaling FFAG, most of the bending occurs in the defocusing magnets. As a result, for an equivalent energy range, magnet apertures in a non-scaling FFAG can be reduced compared with a scaling device. Furthermore, at least for high energies, the ring can be made isochronous at a single energy within the energy range of the machine. This is shown in Fig. 51, which gives the time-of-flight dependence on energy in a typical linear non-scaling FFAG cell. The relatively small time-of-flight variation with energy in these machines allows the use of relatively high frequency RF, such as the 201 MHz RF that is used in the bunching, phase rotation, and cooling channels. This permits reasonably high accelerating gradients.

Linear non-scaling FFAGs become more efficient at higher energies [59], as it is possible to make more passes through the cavities. As a result, our preference is to use FFAGs only at the higher energies, where they become more efficient than RLAs.

Based on our present studies, it appears that a factor of two energy gain is roughly the optimal acceleration range for a single FFAG stage; aperture requirements and time-of-flight range increase very rapidly beyond that. The primary difficulty with linear non-scaling FFAGs is that the time-of-flight depends on transverse amplitude [60], as shown in Fig. 52. Particles with different transverse amplitudes are guided through different regions of longitudinal phase space, as shown in Fig. 53. We see that there is only a limited region of initial phase for which particles with both low and high amplitudes will be accelerated. Once particles reach final energy, low and high amplitude particles will have different phases, since the particles follow trajectories that are roughly parallel to the separatrices. In particular, large amplitude particles arrive later in RF phase than do low amplitude particles. This becomes problematic with multiple FFAG stages, since large amplitude particles should arrive earlier, not later, than low amplitude particles for optimal transmission in the next stage.

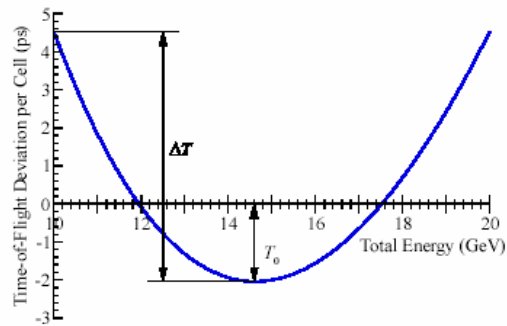

Fig. 51. Time-of-flight as a function of energy in a linear, non-scaling FFAG cell.

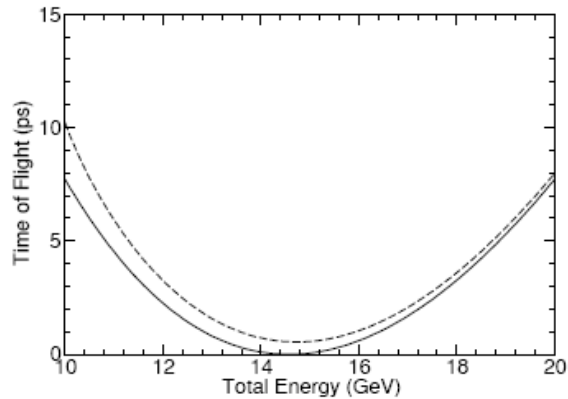

Fig. 52. Time-of-flight as a function of energy for zero transverse amplitude (solid) and 30 $\pi$ mm-rad (dashed) normalized transverse amplitude (twice the transverse action, normalized).

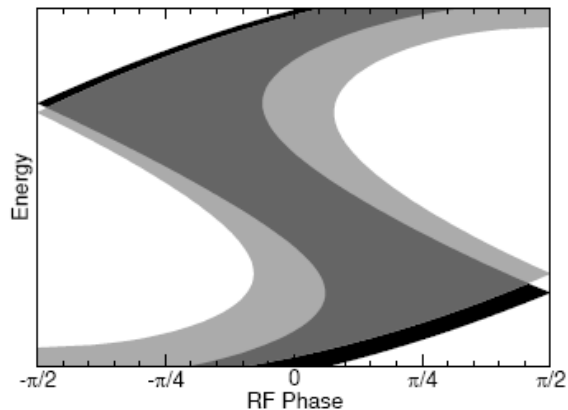

Fig. 53. Phase space channel for low amplitude particles (black) and high amplitude particles (light grey). The overlap region is dark grey. Particles start at low energy (bottom) and are accelerated to high energy (top).

Despite this shortcoming, we believe linear non-scaling FFAGs to be the best option for accelerating to the highest energies. Improvements that help to address the time-of-flight dependence on transverse amplitude are being examined and will be included in future designs. Preliminary simulations [61] suggest that two FFAG stages should result in a tolerable level of longitudinal emittance dilution. With this in mind, we choose (see Fig. 36) to use a non-scaling FFAG to reach 25 GeV, with a second stage being used to reach 50 GeV if required.

### 5.4.3 Isochronous Non-Scaling FFAGs

Aside from the dependence of the time-of-flight on transverse amplitude, the primary factor that controls the number of passes that can be made in an FFAG is the variation of the time-of-flight with energy. Indeed, if the time-of-flight were independent of energy, an arbitrary number of passes through the FFAG should be possible. In practice, of course, the time spent in the FFAG is ultimately limited by decay losses, so the maximum number of passes is still limited.

An 8–20 GeV FFAG was designed that was significantly more isochronous than a "standard" linear non-scaling FFAG would be [2, 62]. To accomplish this, magnetic fields that were highly nonlinear [65] were utilized. Unfortunately, the resulting dynamic aperture was insufficient to accelerate a Neutrino Factory muon beam [63, 64]. The lattice was very sensitive to the precise shape of the magnet end fields; this likely arose from the high degree of nonlinearity in the lattice. Moreover, the rapid field variations required could potentially create difficulties in constructing the magnets. Previous attempts to design a non-scaling FFAG lattice with highly nonlinear magnets have also resulted in insufficient dynamic aperture for accelerating muons in a Neutrino Factory [65, 66]. Therefore, isochronous non-scaling FFAG lattices have not been pursued for accelerating muons.

### 5.4.4 Time-of-Flight Dependence on Transverse Amplitude

It is essential that particles arrive at the appropriate phase of the RF for them to be accelerated and to maintain the appropriate shape of the longitudinal distribution. Usually, an accelerator can be designed assuming that the transverse amplitude has a negligible effect on the longitudinal motion. Unfortunately, this approximation cannot be made for the acceleration of muons due to their large transverse amplitudes. The underlying reason for this is shown schematically in Fig. 54—particles with larger transverse amplitudes have geometrically longer path lengths due to the nonzero angles they make with respect to the reference orbit. As shown in [67], the variation of the time-of-flight with transverse amplitude over a given length of beam line is related to the derivative of the tune $Q$ over that same length of beam line with respect to energy (the chromaticity) by

$$\Delta T = -2\pi (\partial_E Q) \cdot \boldsymbol{J}_n, \tag{10}$$

where $\boldsymbol{J}_n$ is the normalized transverse action (in eV-s). Thus, there is no effect expected if the chromaticity is corrected in the machine. There are, however, two locations where the chromaticity cannot be fully corrected—in the initial accelerating linac, and in the linear non-scaling FFAGs.

For a linac where the magnetic fields are adjusted to keep the phase advance per cell a constant, and where the cell lengths and accelerating gradients are constant, the time-of-flight difference along the full length of the linac can be written as

$$\Delta T = -2\pi \frac{\boldsymbol{\xi} \cdot \boldsymbol{J}_n}{\Delta E} \ln \frac{p_f}{p_i}, \tag{11}$$

where $\xi$ is the chromaticity per cell, defined such that tune is $Q + \xi\delta$, with $\delta$ being the relative momentum deviation, $\Delta E$ being the energy gain per cell, $p_i$ being the reference momentum at the beginning of the linac, and $p_f$ being the reference momentum at the end of the linac. Synchrotron oscillations cause the late arriving particles to be exchanged with earlier arriving ones, mitigating this effect (actually resulting in a shift in the equilibrium energy).

For the last section of linac from Study 2a [21, 41], the resulting additional phase slip at the end of the linac is about 30° at the maximum transverse acceptance. In that section of linac, synchrotron oscillations are negligible. When combined with the earlier sections of linac (which still have a relatively small amount of synchrotron oscillation) and the subsequent linac in the RLA stage (which simply acts like an additional section of the first linac), this amount of phase slip is a potential cause for concern. The complete system has yet to be simulated and analyzed, but our preliminary calculations indicated that it would be preferable to lower the final energy of the initial linac from the Study 2a value of 1.5 GeV to 0.9 GeV.

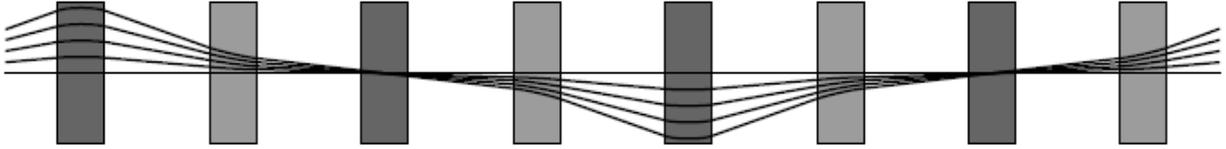

Fig. 54. Trajectories in a FODO lattice, demonstrating that path lengths are longer for particles with larger transverse amplitude.

The same problem occurs in linear non-scaling FFAGs. In the context of the present issue, a linear non-scaling FFAG can be treated as accelerating uniformly and having a tune per cell that does not vary with position in the machine (but does vary with energy). The time-of-flight difference after accelerating through a linear non-scaling FFAG is thus given approximately by

$$\Delta T = -2\pi \frac{\Delta Q \cdot J_n}{\Delta E}, \qquad (12)$$

where $\Delta Q$ is the difference in the tune per cell between the initial and final energies, and $\Delta E$ is the energy gain per cell. This becomes particularly problematic when transferring a beam from one FFAG to another. To result in acceleration, the phase space dynamics in the FFAG require that a particle with high amplitude arrive *earlier* than a low amplitude particle to be accelerated, whereas Eq. 12 indicates that particles with higher transverse amplitude exit the FFAG *later* than those with low transverse amplitude.[8]

### 5.4.5 Addressing the Problem in FFAGs

While we know of no way to completely eliminate the problems caused by the time-of-flight dependence on transverse amplitude in linear non-scaling FFAGs, we can mitigate the emittance growth that this causes by a number of methods:

- We will make optimal use of the longitudinal phase space. The choice of the initial conditions in longitudinal phase space, both for the centroid of the beam and the orientation of the beam ellipse, are important for minimizing the longitudinal distortion [70]. A criterion on the acceptable level of longitudinal distortion will then help determine some of the machine parameters. How to choose these initial conditions and machine parameters has thus far been studied only for the case where the time-of-flight is a symmetric parabolic function of energy [68]. It remains to work out how to optimally choose the initial conditions and machine parameters for the more general case. After doing so, we will use the results for the time-of-flight curves over the entire desired range of transverse amplitudes to optimally choose initial conditions and machine parameters for the real system.

---

[8] Scaling FFAGs do not suffer from this problem, since they have no tune variation with energy. This is part of the motivation to examine an alternative configuration using a scaling FFAG at lower energies.

- We will not force the design to accelerate particles that simultaneously have large transverse and longitudinal amplitudes. This corresponds to transporting a six-dimensional ellipsoidal phase space, rather than a phase space that is a tensor product of a four-dimensional transverse ellipsoid and a longitudinal ellipse. This limitation is unlikely to have major consequences, since we expect it to eliminate only a relatively small number of particles. (It is likely that some other system would cause these particles to be lost in any case.)
- We will add sextupoles to reduce the range of tunes in the machine. Equation (12) shows that, for a uniform acceleration rate (a good approximation for these purposes), the time-of-flight variation with transverse amplitude is proportional to the change in tune from the beginning to the end of the acceleration cycle. Including sextupoles in the lattice reduces this tune difference. At some energies, the chromaticity actually increases locally, so one must be careful about what this does to the longitudinal phase space, but its average over the energy range of the machine is reduced. Unfortunately, the addition of sextupoles reduces the dynamic aperture. The dynamic aperture appears to be acceptable with $\Delta Q$ reduced by 20–30%. However, this requires that the low energy horizontal tune be kept below 1/3 to avoid the $3Q_x = 1$ resonance line, which is driven by the sextupole components, and it may require the avoidance of the $4Q_x = 1$ and $4Q_y = 1$ resonances as well. All of this leads to a reduction in the number of passes the beam can make through the RF [69].
- We will increase the average RF voltage per cell. Equation (12) shows that the time-of-flight variation with transverse amplitude is inversely proportional to the energy gain per lattice cell. Before the discovery of this effect, the optimization procedure [70] generally left a large number of cells in the lattice without RF, since that reduced the magnet aperture and increased the number of passes that were made through the RF with only a modest decay cost. Updated designs will fill these empty cells with RF. If necessary, one can increase the number of RF cells per cavity to two and even further increase the energy gain per lattice cell. This modification would reduce the required number of passes through the RF. It is important to achieve the maximum RF gradient possible in the cavities, and thus R&D in this area is important.
- We will add higher harmonic RF cavities. The shape of the longitudinal phase space (see Fig. 53) causes particles with a different time-of-flight profile *vs.* energy to arrive at the extraction point with different energies as well as different times. Higher harmonic RF cavities will reduce this energy variation.
- We will consider "over-correcting" the chromaticity in the transfer lines. If the chromaticity is made positive in the transfer lines, Eq. (10) indicates that it will pre- or post-correct the time-of-flight variation with transverse amplitude, reducing the average effect, and possibly improving the phase space transmission through the FFAG in the case of pre-correction (see Fig. 53 and the discussion above). It must be ascertained whether over-correcting the chromaticity in the transfer lines has a deleterious effect on beam transmission or emittance growth in those transfer lines.

### 5.4.6 FFAG Tracking Results with Magnet Errors

A non-scaling FFAG lattice has a high superperiodicity—the superperiod is a single cell. If the cell tune is chosen below 0.5, a perfect lattice is free from integer and half-integer resonances. However, once the periodicity is broken due to either misalignment or gradient errors, we must take into account the whole tune of the ring, as opposed to the cell tune. In fact, both horizontal and vertical tunes vary by several units during acceleration. Crossing of integer resonances (driven by alignment errors) and half-integer resonances (driven by gradient errors) becomes a potential concern.

Although the lattice consists only of dipole and quadrupole magnets, the end fields have nonlinearities that can excite nonlinear resonances. The large muon beam normalized acceptance of 30 $\pi$ mm-rad makes kinematic terms non-negligible and they become another source of nonlinear resonances. Breaking of lattice symmetry may enhance the harmonic content of those nonlinear resonances. To assess such effects, a tracking study was performed with various random errors on a 10–20 GeV linear non-scaling FFAG ring. We included alignment and gradient errors. We assumed that the defocusing and focusing quadrupole in each cell were on a single support table and thus could be misaligned together in both the horizontal and vertical directions. We have not yet introduced magnet tilts or longitudinal displacements, as these effects are typically unimportant. Distributions of both alignment and gradient errors are Gaussian with a cut at $2\sigma$. To have decent statistics, 40 different seeds are examined for each case.

Particles are assumed to gain the same amount of energy per cell, independent of the RF phase, to avoid difficulties related to the time-of-flight variation with transverse amplitude. This is necessary to separate particle loss due to resonance crossing from other sources of loss. Acceleration from 10 to 20 GeV takes 17 turns, giving almost the same crossing speed at each individual resonance as would the nominal acceleration. The 500 macroparticles are distributed in an ellipsoidal volume in 4D transverse space with a waterbag distribution function. There is no momentum or phase spread initially.

Although we introduce errors in the lattice, we still assume the distortion of lattice beta functions is small. In other words, the initial particle distribution is matched to unperturbed lattice functions, not to perturbed ones. To ensure this assumption is valid, we chose the initial tune of the ring away from integer and half-integer resonances at the injection momentum. Thus, the initial momentum was changed to 9.965 GeV/c, rather than adjusting quadrupole strengths. In order to judge particle loss, particle amplitude—or more specifically single-particle emittance—is calculated at every cell for all macroparticles. If the amplitude is more than 45 $\pi$ mm-rad, that is, 1.5 times the acceptance, the particle is considered lost.

Figure 55 shows the tracking results for various gradient errors and Fig. 56 shows the tracking results for various alignment errors. In each case, the number of surviving macroparticles is plotted as a function of the errors. Each point corresponds to an individual lattice with different error seed. These figures show that alignment errors should be kept below 100 μm rms and gradient errors should be below $1 \times 10^{-3}$ rms if the maximum allowable loss is taken to be 10%. Separate tracking results confirm that the two errors have no correlation. These tolerances are not trivial to achieve, but are certainly possible with modern technology.

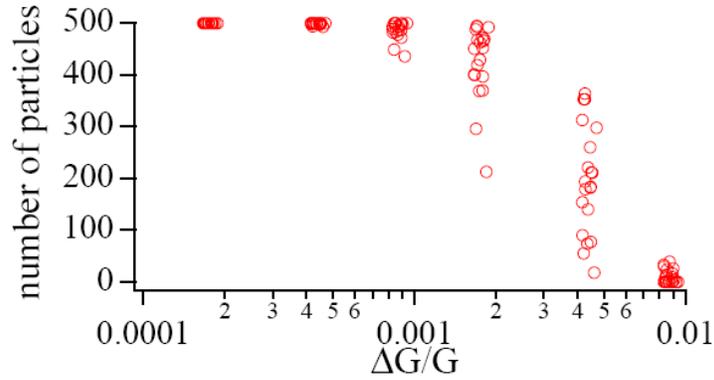

Fig. 55. Number of surviving particles as a function of gradient error.

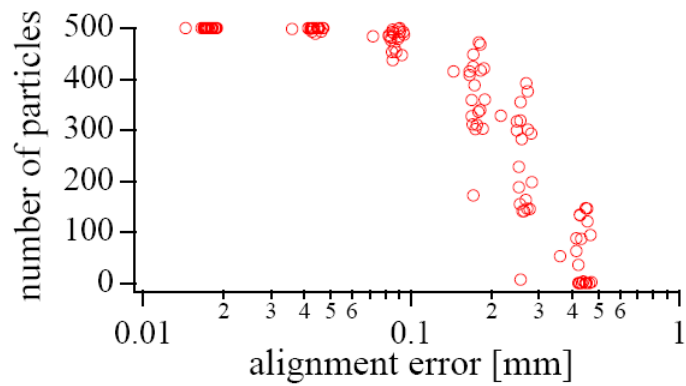

Fig. 56. Number of surviving particles as a function of alignment error.

Further studies will be performed to better understand these error tolerances. We must consider how much emittance growth we are willing to tolerate and adjust the amplitude at which we throw out particles in the simulation accordingly. This may require that apertures be enlarged, and the cost and practicality of doing so must be studied. Closed-orbit distortion is not expected to be a significant problem due to the rapid acceleration. The dependence of particle loss on resonance crossing speed will also be studied. Although there may not be much room for adjustment in reality, higher crossing speeds (i.e., faster acceleration rates) should reduce particle loss. Finally, losses due to individual resonance crossings, as opposed to losses during the entire cycle, should be investigated. Although a beam crosses many resonances, there may only be a few that really affect it. A study of individual crossings, and the relation between particle loss and the driving term of each resonance, would give us more insight into the loss process, and hopefully suggest a way of correction.

### 5.5 Multiple Bunch Trains

As described elsewhere in this document, it is desirable to have multiple bunches accelerated in each proton driver cycle. If those bunches are delivered to the muon cooling and acceleration systems separated by a time greater than the cavity fill time, it would substantially increase the average power required for a muon accelerator, since most of the stored energy in the cavities is thrown away after each muon pulse train. Furthermore, it would increase the average power requirement on the RF power source, since its duty factor would increase substantially.

Instead, one can deliver the bunch trains in rapid succession to the muon cooling and acceleration system. In doing so, care must be taken with the acceleration systems, where multiple passes are made through the RF cavities. This is particularly true for the FFAGs, since the beam extracts a significant fraction of the cavity stored energy in that case. If that stored energy is not promptly restored, different bunch trains will be accelerated by different amounts.

As an example, consider a 10–20 GeV FFAG containing 54 cavities. We assume that the maximum power delivered to the cavity, due to limitations of the input coupler, is 1 MW, and that there is a 4 MW, 24 GeV proton driver delivering some number of proton bunches in rapid succession every 20 ms. Based on our design studies, we take 0.17 muons per proton to be available at the acceleration system. Then, the minimum time between bunch trains and the minimum time for the entire sequence of trains, as a function of the number of trains, is given in Table 16. For a mercury-jet target, if the time between proton pulses hitting the target is too long, the jet will begin to break up and will not provide a suitable target for the later proton bunches. Times indicated in Table 16 for the entire bunch train are, unfortunately, substantially longer than what that the breakup time is assumed to be ($\approx 50\ \mu s$).

Reducing the energy gain per cavity would reduce the time required for the bunch train, but has several drawbacks—it would require substantially more cavities, it would increase the effect of beam loading on different bunches within the same train, and it would worsen the effects of the variation of the time-of-flight with transverse amplitude. Instead, our proposed solution is to drive different cavities with slightly different frequencies, creating a beat wave with a period that is long compared with the time for the bunch train. The phasing of that beat wave can be chosen, in combination with the energy loss from beam loading, to make all the bunch trains gain the same amount of energy. This will, of course, require installing a larger number of RF cavities.

If the proposed approach is effective, then the time between bunch trains can be reduced to any value that is greater than the time spent in a single stage of acceleration. In earlier FFAG designs, the time spent in the 10–20 GeV FFAG stage was about 21 $\mu$s, meaning that if all the bunch trains could take at most 50 $\mu$s, only 3 bunch trains per proton driver cycle would be permissible. Updated FFAG designs may have somewhat smaller times spent in them, since the issue of the time-of-flight dependence on transverse amplitude will be addressed by having a larger average accelerating gradient. Although the beat-frequency solution will reduce the average accelerating gradient, it still appears to be a workable approach to mitigate beam loading effects.

Table 16. Bunch train scenarios for acceleration system.

| Bunches in Train | Time between bunch trains ($\mu$s) | Time for all bunch trains ($\mu$s) |
|---|---|---|
| 2 | 105 | 105 |
| 3 | 70 | 140 |
| 4 | 52 | 157 |
| 5 | 42 | 168 |

# 6. Decay Ring

Conceptual designs have been obtained for both racetrack and triangular shaped, $\mu^+$ and $\mu^-$ decay rings. A 20 (upgradeable to 50) GeV energy has been considered [71], with neutrino detectors at distances of 7500 and ~3500 km. For these baseline distances, racetrack designs need the ring planes tilted downwards by ~36° and ~18°, respectively. A triangle design needs side-by-side rings in a (near) vertical plane, with detectors in (nearly) opposite directions from the rings, in gnomonic projection. If suitable detector sites are available, the triangle rings are favored, as their ~40% greater production efficiencies make them the more cost effective. If suitable sites are not available, the use of racetrack rings in separate tunnels provides the better solution. In the absence of specific sites for the accelerator and detector, we have adopted the more flexible racetrack scenario as our baseline. Both designs are compatible with the Neutrino Factory's pattern of three or five bunch trains, as described in Section 2.3.

Recently, it has been recognized that a bow-tie shape of decay ring has several advantages if neutrino detectors are at 7500 and ~3500 km distances, as specified in the study. Unfortunately, a bow-tie shape preserves the muon polarization, and interferes with the accuracy of the related beam instrumentation [30]. A possible scheme to overcome this drawback is being considered.

## *6.1 Ring Features*

As noted, the use of a single racetrack ring in each of two separately oriented tunnels is proposed as the baseline design[9], as this facilitates finding suitable detector sites. There are two operational possibilities:

1. the $\mu^+$ and the $\mu^-$ bunch trains are injected into separate racetrack rings, with each ring aligned to its own detector
2. each racetrack ring has counter-rotating $\mu^+$ and $\mu^-$ beams

In option 1, which is simpler in terms of injection and transport, at any given time one detector site would look at decays from a single sign of muon. The two beams would be switched periodically by reversing all magnet polarities in both rings. In option 2, where each ring would store counter-rotating beams of both signs simultaneously, each detector would see decays from both $\mu^+$ and $\mu^-$ bunch trains, interleaved in time and separated by about 100 ns.[10] In either option, stored muons decay to neutrinos, which pass from the ring straight sections to the detectors. Depending on whether option 1 or option 2 is chosen, the racetracks have either one neutrino production region, or two of a slightly lower individual efficiency. Although the details have not been worked out, it would be advantageous in option 2 to inject in a utility straight section in the middle of the upper arc rather than in a long straight section.

In the backup scenario, each triangle ring would have two downward sloping production regions, with the $\mu^+$ trains in one ring interleaved in time with the $\mu^-$ trains in the other. Each detector site would accept neutrinos from both rings. Baseline distances needed are 7500 km for one detector, and 2500–3500 km for the other. For these distances, the smallest triangle apex angle (and the best production efficiency) is about 50°, when detectors are in opposite directions (in a gnomonic projection) from vertically aligned rings. Some incline of the rings from an exact vertical plane is expected for most pairs of suitable detector sites.

---

[9] The alternative choice, using two separate isosceles triangle shaped rings in a common, larger tunnel is attractive only if the two detector sites are suitably oriented. We will continue to examine this as a backup option.
[10] The advantage of option 2 is that a single experiment can make use of the full beam intensity if the other detector site (or decay ring) is unavailable.

## *6.2 Ring Specifications*

Detector locations for accelerator sites have to be defined before final parameters can be set. Specifications depend on input beam parameters along with detector and bunch pattern requirements. A compatible set of rings (with the booster orbit half the length of the driver injection orbit, and the proton rotation period in the driver at 10 GeV half that for the muons in the decay rings) are a booster of circumference 400.8 m, an NFFAG with injection (ejection), orbit lengths 801.6 (801.4) m, and a 25 GeV decay ring of circumference 1608.8 m.

An important parameter for the production straights is the ratio of the muon rms divergence angles to the rms opening angles of the decay neutrinos. The ratio has to be ≤ 0.14 at both 20 and 40–50 GeV, for the normalized rms transverse input $\mu^{\pm}$ beam emittances of 4.8 $\pi$ mm-rad. The decay ring apertures are set a factor 50% larger than the beam envelopes, to allow the use of muon beam loss collectors, which are needed due to the megawatt muon beam power involved.

## *6.3 Lattice Designs*

Both the racetrack and triangle rings use bend magnets at the ends of the production straights to separate off neutrinos that arise from muons with large divergence angles. These magnets complicate the lattice designs by creating dispersion in the matching sections to the main arcs. Six-parameter matching is needed, with the dispersion kept small in the regions of large betatron amplitudes. The lattices for the racetrack and triangle rings have different designs for the arcs, the production straights, and the matching sections, though the designs are interchangeable.

### 6.3.1 Racetrack Ring

A layout of a 1608.8 m circumference racetrack ring is shown in Fig. 57. There are two arcs, each with 15 FODO cells of superconducting dipole and quadrupole magnets. If a single neutrino production straight section is used, it has a length of 600.2 m and a production efficiency of 37.3%. The other long straight section has the collimators, tune control, and RF systems. If a counter-rotating, $\mu^{\pm}$ beam option is used, the lattice is modified for two shorter production straights of a slightly reduced efficiency.

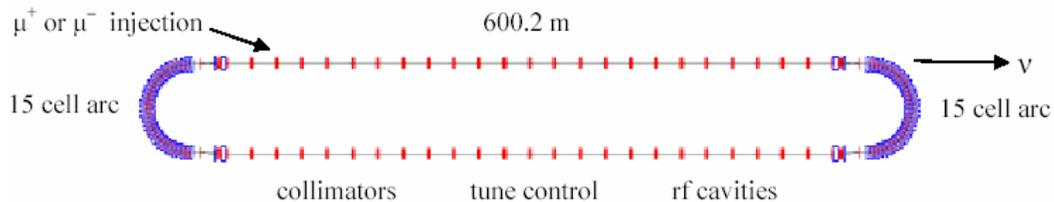

Fig. 57. Schematic layout for a racetrack shaped, muon decay ring.

Separate tunnels are used for the two racetrack rings as they are aligned to different detectors. Ring planes are sloped downwards at an angle of ~ arcsin ($L/2R$), where $L$ is the distance to the detector and $R$ is the equatorial radius. As noted, for the distances proposed of 7500 and 3500 km, the tilt angles are ~36° and ~18°, and the maximum depths of the tunnels are ~435 m and ~185 m, respectively. The width of the tunnels is not as great as that of the common tunnel used for the triangle rings. The $\mu^+$ and $\mu^-$ beam lines from the 20 (50) GeV, $\mu^\pm$ accelerating ring have branches passing to each racetrack, for a total of four beam line tunnels. Services and service buildings have to be provided for both rings and all beam line tunnels.

The racetrack decay rings are based on a design concept from an earlier Neutrino Factory study [40], though parameters and some ring elements have changed. The beta function values in the production straight section (see Fig. 58) are reduced to ~153.0 m, while the transverse acceptance of the ring is increased to 67.5 $\pi$ mm-rad. The dispersion introduced by the dipoles at the ends of the production straight section increases throughout the arc matching sections until a six-parameter match to the arcs is obtained.

For the proposed racetracks, the ratio of the muon rms divergence angle to the rms opening angle of the decay neutrinos is ~0.11 at 20 GeV, assuming a normalized rms muon emittance of 4.8 $\pi$ mm-rad. If the upgrade lattice is unchanged, the ratio scales with $\sqrt{\gamma}$ at higher energy, becoming ~0.17 at 50 GeV.

### 6.3.2 Triangle Ring

A layout drawing of an isosceles triangle ring, with a 1608.8 m circumference and a 52.8°, apex angle, is shown in Fig. 59. Two 398.5 m long downward sloping production straights give a neutrino production ratio efficiency of 2 × 24.8%. A maximum efficiency results when the apex angle is minimum, with the detectors in opposite directions (in gnomonic projection) from two vertically aligned rings in the same tunnel. When the detector sites are not opposite, it is necessary to tilt the plane of the rings about a production straight axis, and increase the apex angle until the straights and detectors are again re-aligned.

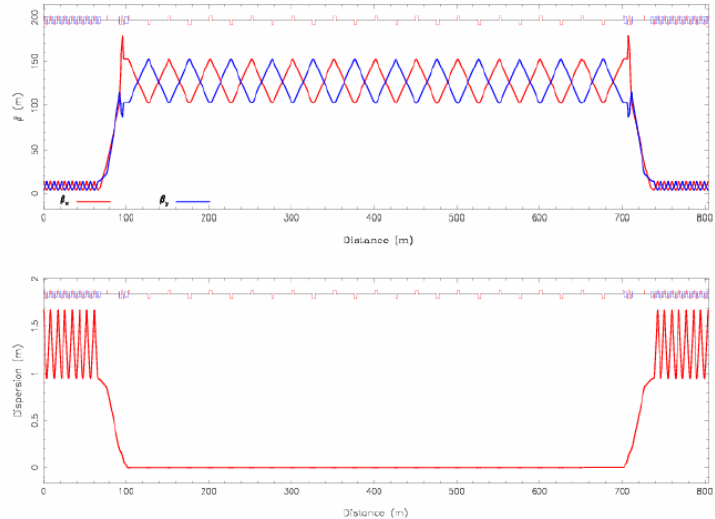

Fig. 58. Betatron and dispersion functions in a racetrack ring.

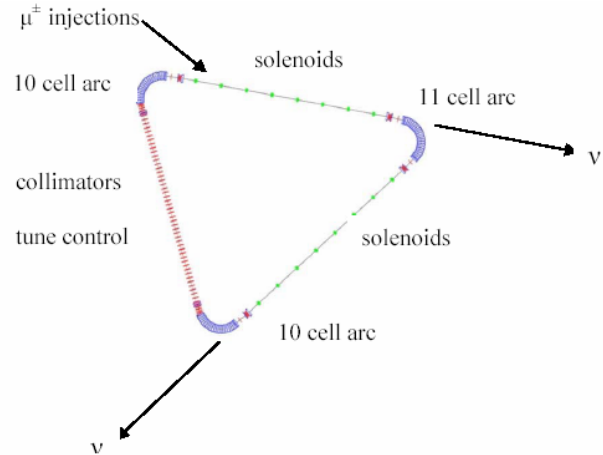

Fig. 59. Schematic layout for a 52.8° apex angle, isosceles triangle, muon decay ring.

The arc cells have a FODO design, and the maximum values for betatron and dispersion functions (see Fig. 60) are $\beta_y$ = 12.67 m, $\beta_x$ = 12.67 m and $D_x$ = 1.44 m. At the ends of the arcs, the dispersion function is $D_x$ = 0.85 m. There are 11 cells in the triangle apex and 10 cells in each of the other arcs. Each arc cell provides 11.1° of bending, for a total of 344.1°. (The remaining 15.9° of bending is distributed among the matching cells between the arcs and the production straight sections and those needed between the arcs and the beam loss collimators.) The arc cells are 8.2 m in length and consist of a pair of 2.9-m superconducting combined-function magnets (one focusing, one defocusing), separated by 1.2 m drift sections. Sextupole components are included in the combined-function magnets to reduce the lattice chromaticity. The betatron phase advance per cell is 72°, so that first- and second-order sextupole terms cancel over a 5-cell unit.[11] To minimize dispersion leakage, the arcs begin and end with a defocusing magnet.

One noteworthy feature of this design is that focusing in each production straight is provided by eight 4.04-T superconducting solenoids, arranged symmetrically. A figure of merit for the production straight focusing is given by the inverse of the maximum lattice $\beta\gamma$ function, which is ~1 for solenoids, but is $(1 - \sin \mu)/2$ for thin lens FODO cells, where $\mu$ is the half-cell phase advance. For equal muon divergence angles, the maximum $\beta$ value for solenoid lenses is thus about half the value found in FODO focusing cells. This lowers the beam size in the production straights and the adjacent matching sections, which should improve the ring dynamic aperture. The maximum beam diameter in the production straight is 265 mm at 20 GeV, for 30 $\pi$ mm rad normalized transverse emittance and a $\beta$ of 94.3 m at the solenoid focusing waists. Aperture diameters are 50% larger than this (about 398 mm). The ratio of the muon beam rms divergence angles to the rms opening angles of the decay neutrinos is 0.1, assuming a normalized rms emittance of 4.8 $\pi$ mm rad.

There are 35 m long, six-parameter matching sections between the arcs and the production straights. Each of these has a gradient magnet, a dipole magnet and four quadrupoles. Dispersion at the end of the arc is reduced to zero in the matching section. Maximum beta functions in this region are $\beta_y$ = 124 m and $\beta_x$ = 101 m.

---

[11] The magnets in the central cell of the 11-cell apex arc do not contain sextupole components.

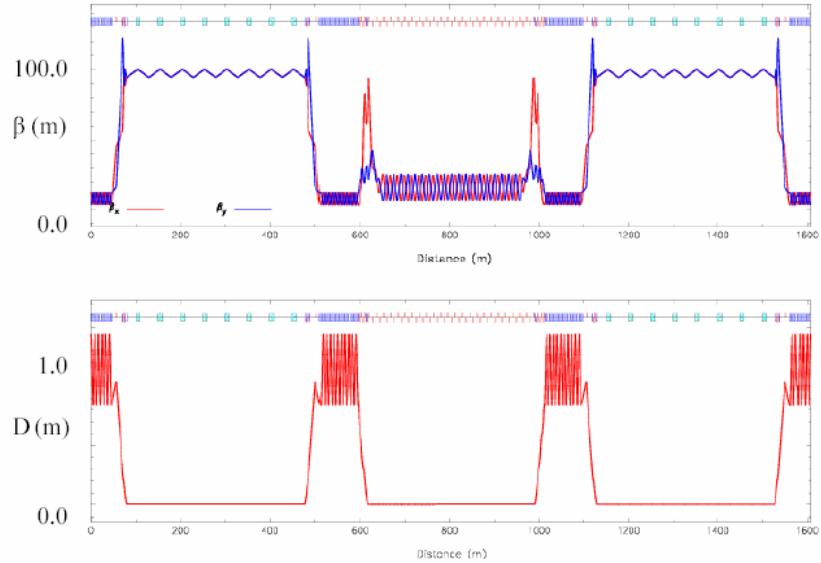

Fig. 60. Lattice betatron and dispersion functions for the 1608.8 m, triangle decay rings.

The vertical straight section opposite the triangle apex is mirror symmetric about its center. There is a zero dispersion central section of FODO cells, where two quadrupole types can be varied to control the ring betatron tunes. For each adjacent section, there are a further five quadrupoles for betatron matching, followed by two dipoles and six quadrupoles that provide a six-parameter match to the neighboring arc. This scheme allows the tune control of the ring to be done independently of the dispersion matching.

The warm bore tubes of the superconducting arc magnets will be clad with lead to absorb the power from $e^{\pm}$ arising from $\mu^{\pm}$ decays. Direct $\mu^{\pm}$ wall losses also cause magnet heating and, to reduce these, there are beam loss collimators in four tune-control cells[12]. Primary transverse collimators are followed by secondary collimators downstream, after betatron phase shifts of 20°, 90°, and 160°.

Bunch trains are injected into the upstream end of the production straights that are nearest to the surface (see Fig. 59). An injection septum magnet is located just downstream of the first solenoid, and 24 m of fast kicker magnets are located between the second and third solenoids. The required kicker rise times are 675 (or 993) ns for 5 (or 3) bunch trains. For five bunch trains, the estimated number of 5 kA, 50 kV systems per ring for the multi-pulsing of the kickers is 14, together with 2 spares. The multi-pulsing feature is an item requiring R&D.

Lattice modifications are planned when the rings are upgraded from 20 to 40 or 50 GeV. Some matching components would be repositioned and the solenoid focusing would be weaker[13] (though the fields are increased). The two rings would require a realignment as a result of the modifications, and the fields at the central orbits of the superconducting arc magnets must be increased to ~ 5.6 T at 50 GeV.

---

[12] These cells utilize conventional rather than superconducting quadrupoles to avoid quench problems.
[13] Beta values at the focusing waists of the solenoids increase from 94.3 to 160.3 m in order to limit the increase in the ratio of muon-to-neutrino divergence angles. With the change in beta values, the divergence ratio only increases from 0.1 at 20 GeV to 0.12 at 50 GeV; without the changes, it would increase to 0.16.

For both the racetrack and triangle rings a full study of dynamic aperture (both on- and off-momentum) is called for. This study must include the effects of all error sources, including magnet strength and field errors, and alignment errors. Compared with most rings, a muon decay ring has the advantage of needing to store the beam for only about 500 turns, so our expectation is that ring performance should not be overly sensitive to machine errors. Detailed studies of the baseline and alternative rings will be carried out as part of the upcoming IDS-NF study.

### *6.4 RF and Diagnostic Systems*

The injected beam trains consist of 201.25 MHz, $\mu^+$ or $\mu^-$ bunches with a momentum spread range, $|\Delta p/p|$ of 0.01–0.03. The debunching rate without any ring RF systems is $dT/dt = \eta \, \Delta p/p$, where $2T$ is the bunch time duration and $\eta = \gamma_t^{-2} - \gamma^{-2}$ is the phase-slip factor. For 20 GeV muons in the triangle ring, the rate is 0.5 ns per turn for $|\Delta p/p|$ of 0.02. Thus, adjacent bunches merge in the rings after five turns, and for the five (or three) interleaved bunch trains, the time gaps decrease from 139 (or 298) ns initially to the minimum specified gap of 100 ns in 76 (or 388) turns. An RF system is clearly needed for five-bunch trains, but may not be essential for three-bunch trains. In the racetrack rings, an RF system is needed for the option 2 case with counter-rotating bunch trains, but not otherwise.

Even if not needed for keeping the muon bunch trains separated, an RF system has benefits for diagnostic systems and for reducing $\Delta p/p$. The bunch frequency is superior to the bunch train frequency for typical beam position monitor designs, and a common 201.25-MHz beam position monitor design for all muon rings is an attractive option. Bunch lengths may be measured during the initial bunch rotation to $\Delta\varphi_{RF} = \pm\pi/2$, as the $\Delta p/p$ reduces in decreasing RF fields.

Spin depolarization measurements, as the beam debunches with the RF off, have been proposed for calibration of the muon energy [30]. As RF is needed for some operations, an RF bunch rotation from small initial $\Delta\varphi$ to $\pm \pi/2$ while at an intrinsic resonance is considered for the depolarization. The $n - Q_y = (g - 2)\gamma/2$ resonance occurs at 20 GeV for a harmonic $n$ of 13, a $Q_y$ of 12.78 and a spin tune of 0.222.

### *6.5 Summary*

Conceptual designs have been obtained for racetrack- and triangular-shaped $\mu^+$ and $\mu^-$ decay rings. A 20 GeV (upgradeable to 50 GeV) energy has been considered, with neutrino detectors at distances of 7500 and ~3500 km. Racetrack designs need the ring planes tilted downwards by ~36° and ~18°. Triangle designs need side-by-side rings in a (near) vertical plane, with detectors in (nearly) opposite directions from the rings, in gnomonic projection. In the absence of specific sites, the use of the racetrack rings in separate tunnels provides the more flexible solution and was adopted as the baseline scenario. However, if suitable detector sites were available, the triangle rings would be favored, as their ~40% greater production efficiencies make them the more cost effective solution.

# 7. R&D Needs

R&D activities in support of a Neutrino Factory have been ongoing for many years. In recent years, the effort has become more and more a coordinated international effort. In what follows we will highlight some of the main R&D issues that need to be studied in preparation for beginning facility construction.

## 7.1 Proton Driver

In Section 2, we described a nonlinear non-scaling FFAG ring that could serve as a proton driver ring. As this concept is new and untried, the fabrication and testing of a low-energy electron model is called for. Continued development and testing of tracking codes adequate for this parameter regime are also important. For example, space-charge issues will be significant. Development of collimators to protect key machine components is needed. Both primary beam loss and beam halo need to be considered.

In the case of the RF systems, development efforts aimed at improved designs for low-frequency high-gradient cavities must continue. Because the beam power is high, beam loading is a matter for concern and must be studied both computationally and, ultimately, experimentally.

For the linac-based designs, details of the ancillary rings (accumulator and compressor) need to be specified. To permit hands-on maintenance, beam losses must be kept to a minimum. This will involve careful studies of vacuum issues, instabilities, and beam halo formation. Both J-PARC and LHC are developing the tools for this, and participation in such activities will be a help for proton driver development.

## 7.2 Target

Work on the liquid-Hg jet target is well along in the context of the MERIT experiment. This work needs to be completed and analyzed with high priority. Determination of acceptable single pulse and pulse train durations must be made. As a possible follow-on, it will be worth exploring other high-$Z$ targets that are not liquid at room temperature but have a low melting point, e.g., a Pb-Bi eutectic. Much of this can be done off-line with a modified MERIT apparatus. An assessment of the possible need for beam tests should be part of this program.

Solid targets remain a possibility for a Neutrino Factory, at least at the 1 MW proton driver level. Tests to determine the power-handling capability of solid materials should continue in order to identify the practical limits of this technology. This will involve shock tests and irradiation studies to understand the changes in materials properties in the Neutrino Factory target environment. Determining acceptable single-bunch and bunch-train spacing parameters is necessary for solid targets also. Development of one or more practical implementation options for solid targets is needed. A beam test of such a system is also highly desirable.

At present, our information on pion production rates and their dependence on the proton beam parameters (particularly bunch length and energy) comes solely from model calculations. Incorporating measurements of production rates into our performance estimates will be critical in deciding on the optimum parameters of the proton driver.

### *7.3 Front End*

Foremost in this area is the demonstration of ionization cooling that will take place in the MICE experiment. This will take several more years to complete. Developing the various components needed remains a high priority. In particular, high-gradient RF cavities that operate in a strong solenoidal field are needed. Both vacuum cavities having irises closed with beryllium disks and $H_2$-gas-filled cavities have been proposed and both need further study. For the vacuum cavity, the primary issue is the observed degradation of gradient in a strong magnetic field. For the gas-filled cavity, the main issue is whether the gas maintains its desirable insulating properties when subjected to an intense beam of ionizing radiation. This test requires an intense beam, but does not require muons. Absorber thermal tests with LiH sandwiched in beryllium must be carried out. As a follow-on to MICE, building and testing a section of "Guggenheim" cooling channel will provide options for producing 6D cooling, thus improving the compatibility between Neutrino Factory and Muon Collider designs.

Experimental studies of muon multiple scattering are being analyzed and should be included in simulations of the cooling process. It is not expected that the cooling performance will be markedly changed by such details, but this needs to be confirmed.

In terms of simulations, we must optimize the machine by balancing the cooling channel performance against the acceptance of the acceleration system. We also need to evaluate the robustness of our technical solutions by means of error studies.

### *7.4 Acceleration*

The primary acceleration system R&D activity will be to participate in the EMMA experiment to test an electron model of a non-scaling FFAG. This type of accelerator is presently untested, and we need to know whether our performance simulations are correct. There are many beam dynamics issues that have arisen during the course of the ISS, most notably the dependence of time-of-flight on transverse amplitude, that need to be fully understood. At present, it appears that no more than two FFAG systems can be cascaded. We need to develop and test mitigation techniques to improve the situation.

Because the acceleration system layout tends to be tightly spaced (both RLAs and FFAGs), we need to demonstrate that we can reliably operate superconducting RF cavities in close proximity to high-field magnets, and that we can achieve the requisite gradients at 201 MHz. Initial work on this at Cornell was encouraging, but much remains to be done.

Another area that needs exploration is the use of high-frequency cavities in a scaling FFAG. If there were intractable issues that arose with non-scaling FFAG designs, the so-called harmonic-number-jump acceleration scheme might be a viable fallback. This needs first to be studied in detail with simulations, but could lead to a hardware test if the calculation results look encouraging. Such an approach, if needed, would be more tractable for an early acceleration stage, where conventional FFAG magnets can be used.

Because we require the largest practical acceptance for the acceleration system, the normal paraxial approximation does not hold. New tracking tools are being developed for this purpose, and these need to be checked carefully with other codes and with experiments whenever possible.

### *7.5 Decay Ring*

The decay ring requires several novel superconducting magnets. They must be combined-function devices and must accommodate the substantial heat load from decay electrons from the muon beam. Designs for these magnets are needed, along with the corresponding cost estimates. Large aperture injection kickers capable of operating at 50 Hz must be developed.

Tracking studies with errors need to be continued, using specialized codes like ZGOUBI [7] that can handle this parameter regime. Polarization studies are needed to see whether a bow-tie ring is a possible configuration for the decay rings.

## 8. Summary

In this document we have summarized the findings of the ISS Accelerator Working Group. We have developed parameters for the proton driver, determined an optimum target implementation, defined a front-end scenario, and proposed a viable acceleration scheme. Several decay ring geometries have been considered and compared. The present baseline assumes a pair of racetrack rings. The alternative triangle geometry, which has somewhat higher efficiency, would be preferred if suitable detector sites are available.